\documentclass[aip,pof,amsmath,amssymb,preprint]{revtex4-1}
\usepackage{verbatim}
\usepackage{graphicx} 
\usepackage{enumitem}
\usepackage{amsmath}
\usepackage{amssymb}
\usepackage{mathrsfs}
\usepackage{tabularx} 
\usepackage{subfigure}
\usepackage{color} % gestion de différentes couleurs
\definecolor{linkcolor}{rgb}{0,0,0.6} % définition de la couleur des liens pdf
\usepackage[ pdftex,colorlinks=true,
pdfstartview=FitV,
linkcolor= linkcolor,
citecolor= linkcolor,
urlcolor= linkcolor,
hyperindex=true,
hyperfigures=false]
{hyperref} % fichiers pdf 'intelligents', avec des liens entre les références, etc.

\draft

\begin{document}

\title{Fluctuations and large deviations of Reynolds stresses in zonal jet dynamics}

\author{F. Bouchet}
\email{freddy.bouchet@ens-lyon.fr}
\affiliation{Univ Lyon, Ens de Lyon, Univ Claude Bernard, CNRS, Laboratoire de Physique, F-69342 Lyon, France}

\author{J. B. Marston}
\email{marston@brown.edu}
\affiliation{Department of Physics, Box 1843, Brown University, Providence, RI 02912-1843 USA}

\author{T. Tangarife}
\affiliation{Univ Lyon, Ens de Lyon, Univ Claude Bernard, CNRS, Laboratoire de Physique, F-69342 Lyon, France}

\date{\today}

\begin{abstract}
The Reynolds stress, or equivalently the average of the momentum flux, is key to understanding the statistical properties of turbulent flows.  Both typical and rare fluctuations of the time averaged momentum flux are needed to fully characterize the slow flow evolution. The fluctuations are described by a large deviation rate function that may be calculated either from numerical simulation, or from theory. We show that, for parameter regimes in which a quasilinear approximation is accurate, the rate function can be found by solving a matrix Riccati equation.  Using this tool we compute for the first time the large deviation rate function for the Reynolds stress of a turbulent flow. We study a barotropic flow on a rotating sphere, and show that the fluctuations are highly non-Gaussian.  This work opens up new perspectives for the study of rare transitions between attractors in turbulent flows. 
\end{abstract}

%\tableofcontents



\maketitle

\section*{Tom\'as}
Regretfully, Tom\'as Tangarife suddenly and unexpectedly passed away a few months before completing the research reported in this paper. Most of the science discussed in this paper was developed in patient work by Tom\'as, and is part of his PhD thesis. F. Bouchet and J. B. Marston pay homage to the unique friendship and passion for science of Tom\'as, and would like to remember the intense and enriching collaboration that led to these scientific results. Tom\'as' quiet and constant character, his generosity, and his deep thoughts, were always a source of happiness and joy to his friends and colleagues.

\section{Introduction}

For a wide range of applications, in physics, engineering, and geophysics,
the determination of the behavior of the average or typical behavior
of a turbulent flow is a key issue.  Since the work of Reynolds more than one
century ago, the role of momentum fluxes and their divergence, or
their averages called Reynolds stresses, have been recognized to play
the key role. In order to be more specific, we now consider the very
simple case of a two dimensional flow on a plane or in a channel, with an average
 flow that is parallel to
the $\mathbf{e}_{x}$ direction, $U(y)\mathbf{e}_{x}$ (where $x$
and $y$ are Cartesian coordinates).  We also assume that all averaged
quantities do not depend on $x$. The spatially averaged equation of motion for the 
fluid reads 
\begin{equation}
\frac{\partial U}{\partial t}=-\frac{\partial}{\partial y}\mathbb{E}\left(<uv>\right)+D\left[U\right],\label{eq:Reynolds}
\end{equation}
where $D[U]$ is the average dissipation operator, $\mathbb{E}\left(<uv>\right)$
is the Reynolds stress, and $\frac{\partial}{\partial y}\mathbb{E}\left(<uv>\right)$
is the momentum flux divergence along the $\mathbf{e}_{x}$ direction.
The symbol $\mathbb{E}$ is either an ensemble
or time average (for a time average $\partial U/\partial t=0$), while $<.>$ denotes a spatial average. 
The spatial average is an average along the $\mathbf{e}_{x}$ direction. 
The spatial average can be avoided, but it is often useful to include for practical reasons. 
Because the Reynolds stress is the key quantity that determines the
average flow behavior it has been extensively studied experimentally,
numerically and theoretically, for a wide range of turbulent flows
(see for instance classical turbulence textbooks \cite{tennekes1972first,pope2001turbulent}.

Beyond the average value, fluctuations
of the momentum flux $<uv>$, or its divergence $\frac{\partial}{\partial y}\left(<uv>\right)$,
are very important quantities in a variety of dynamical circumstances.
By contrast with the average value, as far as we know, no work has
been devoted so far to study such fluctuations, and we undertake this task as the main aim
of the paper.  An important example of when fluctuations play an important role is in the case of time
scale separation between the typical time $\tau_{U}$ for the evolution
of the parallel flow (or jet) and the time $\tau_{e}$ for the
evolution of the turbulent fluctuations (or eddies): $\tau_{e}\ll\tau_{U}$.
Such time scale separation is common when the parallel flow has
a very large amplitude; classical examples include some regimes of
two dimensional, geostrophic, or plasma turbulence.  Then, following the classical results of stochastic averaging for systems with two timescales, a natural
generalization of Reynolds average equation is 
\begin{equation}
\frac{\partial U}{\partial t}=-\frac{\partial}{\partial y}\mathbb{E}_{U}\left(<uv>\right)+\frac{\partial}{\partial y}\zeta_{U}+D\left[U\right],\label{eq:Slow_Stochastic_Dynamics}
\end{equation}
where now $\mathbb{E}_{U}$ means an average over a time window
short compared to the typical time evolution of the parallel flow
$U$, and we still call $\mathbb{E}_{U}\left(<uv>\right)$ the Reynolds
stress that now depends on the state of $U$ at time $t$, and $\zeta_{U}(y,t)$
characterizes the Gaussian typical fluctuations of the momentum flux $<uv>$. $\mathbb{E}_{U}\left(<uv>\right)$ and $\zeta_{U}$ represent two aspects of the action of the unresolved eddies on the mean flow, the average and typical fluctuations respectively. In such
a situation of time scale separation,  $\zeta_{U}$ is a white in time Gaussian field whose variance is related through a Kubo formula to the variance of the time average of the momentum flux
\begin{equation}
r_v=\frac{1}{T}\int\text{dt}\,<uv>,\label{eq:Time_Averaged_Reynold_Stress}
\end{equation}
where the time average is over a time window of duration $T$, which
is assumed to be short compared to the time scale for the evolution
of $U$, but large compared with the evolution of the turbulent fluctuations:
$\tau_{e}\ll T\ll\tau_{U}$. We call the fluctuation of (\ref{eq:Time_Averaged_Reynold_Stress})
the Reynolds stress fluctuations (the fluctuation of the time averaged momentum fluxes, over finite but long times $T$). 

In many instances, rarer and non Gaussian fluctuations are also important. Then \eqref{eq:Slow_Stochastic_Dynamics} does not contain the relevant information and one wants to go beyond the study of the second moment of  \eqref{eq:Time_Averaged_Reynold_Stress}. In the asymptotic regime $\tau_{e}\ll T$, the probability distribution
function of $r_v$ takes a very simple form $P(r_v,T)\underset{T\rightarrow\infty}{\asymp}\exp\left(-TI_v[r_v])\right)$,
where $\asymp$ is a logarithmic equivalence (the logarithms of the
right and left hand sides of the equation are equivalent in the limit $T \rightarrow \infty$).
This relation is called the large deviation principle. (For a review, see Ref. \onlinecite{touchette2009large}.)
The large deviation rate function
$I_v[r_v]$  characterizes the fluctuations of the time averaged Reynolds
stress, both typical (the second variations of $I_v[r_v]$ 
gives the statistics of $\zeta_{U}$), and very rare.  In many examples of turbulent flows, it has
been observed that the dynamics has several "attractors'' (see for instance \cite{Bouchet_Simonnet_2008} and references therein ; by ``attractor'' we mean here stationary solutions of the deterministic Reynolds equation $\frac{\partial U}{\partial t}=-\frac{\partial}{\partial y}\mathbb{E}_{U}\left(uv\right)$). Then rare fluctuations of the Reynolds stress
characterized by the large deviation rate function $I_v$, are responsible
for rare transitions between attractors. For all these reasons, it
is very important to be able able to compute $I_v$ and to be able to
study its properties from a fluid mechanics point of view.\\

We develop theoretical and numerical tools to study
Reynolds stress fluctuations, and compute the large deviation rate
function $I_v$.  First we sample empirically (from time series generated from numerical simulations) the large deviation rate function, using the method developed in reference \onlinecite{rohwer2014convergence}. In addition to this empirical approach, we determine the Reynolds stress fluctuations
and large deviation rate function directly for the case of the quasilinear
approximation to the full non-linear dynamics.  The quasilinear approximation
amounts at neglecting the eddy-eddy interactions (fluctuation + fluctuation $\rightarrow$ fluctuation 
triads) while retaining interactions between the mean flow and
the eddies, and may thus be expected to be accurate when the magnitude of the average flow is much larger than the fluctuations. Such a quasilinear
approximation, investigated at least as early as 1963 by Herring \cite{herring1963investigation},
is believed to be accurate for the 2D Navier-Stokes equation,
barotropic flows, or quasigeostrophic models, on either a plane, a
torus, or a sphere, for a range of parameters (discussed below).  Two dimensional flows
are a particularly favorable setting for the quasi-linear approximation because, as Kraichnan
showed in his seminal 1967 paper \cite{Kraichnan:1967jk}, an inverse cascade of 
energy to the largest scales is expected, leading to the formation of coherent structures
with non-trivial mean flows \cite{Kraichnan:1980uy}.  For unforced perfect flows, the large scale structures 
can be predicted through equilibrium statistical mechanics (see for instance \cite{BouchetVenaille-PhysicsReport}). 
For forced and dissipated flows eddies both sustain, and 
interact with, the large-scale flows, and both processes are captured by the quasi-linear approximation.
By contrast, the scale-by-scale cascade of energy that plays a central role in Kraichnan's picture \cite{Kraichnan:1967jk}
relies on eddy + eddy $\rightarrow$ eddy processes that are neglected in the quasi-linear approximation \cite{farrell2003structural,marston2014direct}.

The quasilinear
approximation has been shown to be self-consistent \cite{Bouchet_Nardini_Tangarife_2013_Kinetic} in the limit when
a time scale separation exists between a typical large scale flow
inertial time scale $\tau_{i}$ and a flow spin up or spin down time
scale $\tau_{s}$: $\tau_{i}\ll\tau_{s}$ (then $\tau_{U}\simeq\tau_{s}$
and $\tau_{e}\simeq\tau_{i}$). This time scale separation condition
may however not be necessary. Other factors may favor the validity
of the quasilinear approximation, for instance the forcing of the
flow through a large number of independent modes, through either a
broad band spectrum, or a small scale forcing, keeping the total energy injection rate fixed. The energy transfer is then the same for all forcing spectrums, but with a braod band spectrum each eddy has reduced amplitude, lessening the
interaction between eddies. The range of validity
of the quasilinear approximation has not been fully understood yet. When the quasilinear approximation is valid, and when one further assumes that the forcing acts on small scales only, one can predict explicitly the averaged Reynolds stress \cite{srinivasan2014reynolds,laurie2014universal,woillez2016computation} and sometimes the averaged velocity profile. The Gaussian fluctuations of the Reynolds stress may be parameterized phenomenologically \cite{farrell2003structural,marston2014direct}. The spatial structure of the Gaussian fluctuations has also been studied theoretically.  It has been proven to have a singular part with white in space correlation function and a smooth part (see \cite{Bouchet_Nardini_Tangarife_2016_kinetic_Zonal_Jets}, section 1.4.3, or \cite{tangarife-these}, see also \cite{BouchetNardiniTangarife2015}).
 
Within the context of the quasilinear approximation, we show that the Reynolds stress fluctuations
and its large deviation rate function can be studied by solving
a matrix Riccati equation. The equation
can be easily implemented and solved by a generalization of the classical
tools used to solve Lyapunov equation for the two-point correlation
functions. This mathematical result is the main reason
why we study the Reynolds stress fluctuations for the quasilinear
dynamics in this first study. Moreover we show that the matrix Riccati equation is a much more computationally 
efficient way to study rare fluctuations than through the traditional route of direct numerical simulation. 
The calculation is illustrated for the
case of barotropic flow on the sphere \cite{marston2014direct}, for which the relevance of
the quasilinear approximation, over certain parameter ranges, has been recognized for a some time now.  
For the case of a barotropic flow it
is technically more convenient to discuss the dynamics in terms of
the equation of motion for the vorticity, so we study the corresponding
Reynolds stress that drives the vorticity. 

Section \ref{sec:Barotropic-equation-and} introduces the barotropic equation on the sphere and its
quasilinear approximation. Section \ref{sec:Equal-time-statistics-of} discusses the fluctuations
of the Reynolds stresses, without time average. Section \ref{sec:large_deviations} is an
introduction to averaging for stochastic processes. It explains pedagogically
how an equation for the slow degrees of freedom, for instance the
Reynolds equation (\ref{eq:Slow_Stochastic_Dynamics}), can be obtained.
The relation between the statistics of the noise term, $\zeta_{U}$,
in equation (\ref{eq:Slow_Stochastic_Dynamics}), and the large deviation
of the Reynolds stress (\ref{eq:Time_Averaged_Reynold_Stress}) is
explained. A short introduction to the large deviation rate function
is also provided. Finally, the matrix Riccati equation that permits direct calculation of the large deviation rate function is derived both in a
general framework, and in the case of the quasilinear approximation
of the barotropic equation on the sphere. Section \ref{sub:Gaussian-approximation-of} uses the solution
of the matrix Riccati equation in order to study numerically the zonal
energy balance and the time scale separation in the inertial limit.
Section \ref{sec:Large_Deviations_Reynolds_stresses} discusses the computation of the large deviation rate
function for the time averaged Reynolds stresses of the barotropic
equation on the sphere. Section \ref{conclusions_perspectives} discusses the main conclusions
and presents some perspectives. 

%%%%%%%%%%%%%%%%%%%%%%%%%%%%%%%%
%%%%%%%%%%%%%%%%%%%%%%%%%%%%%%%%
%%%%%%%%%%%%%%%%%%%%%%%%%%%%%%%%
\section{Barotropic equation and quasi--linear approximation\label{sec:Barotropic-equation-and}}

Here we discuss the barotropic equation and its quasilinear approximation that is expected to be valid when a time scale separation exists between the typical time for the evolution of the zonal flow and that of the evolution of the eddies.  We study the dynamics of zonal jets in the quasi-geostrophic one-layer barotropic model on a sphere of radius $a$, rotating at rate $\Omega$,
\begin{equation}
\left\lbrace \begin{aligned} & \frac{\partial\omega}{\partial t}+J(\psi,\omega)+\frac{2\Omega}{a^{2}}\frac{\partial\psi}{\partial\lambda}=-\kappa\omega-\nu_{n}\left(-\Delta\right)^{n}\omega+\sqrt{\sigma}\eta,\\
 & u=-\frac{1}{a}\frac{\partial\psi}{\partial\phi},\quad v=\frac{1}{a\cos\phi}\frac{\partial\psi}{\partial\lambda},\quad\omega=\Delta\psi
\end{aligned}
\right.\label{eq:barotropic-topography-d}
\end{equation}
where $\omega$ is the relative vorticity, ${\bf v}=(u,v)$ is the
horizontal velocity field, $\psi$ is the stream function and $J(\psi,\omega)=\frac{1}{a^{2}\cos\phi}\left(\partial_{\lambda}\psi\cdot\partial_{\phi}\omega-\partial_{\lambda}\omega\cdot\partial_{\phi}\psi\right)$
is the Jacobian operator. The coordinates are denoted $\left(\lambda,\phi\right)\in[0,2\pi]\times[-\pi/2,\pi/2]$,
$\lambda$ is the longitude and $\phi$ is the latitude. All fields
$\omega,u,v$ and $\psi$ can be decomposed onto the basis of spherical
harmonics $Y_{\ell}^{m}(\phi, \lambda)$, for example 
\begin{equation}
\psi\left(\phi, \lambda \right)=\sum_{\ell=0}^{\infty}\sum_{m=-\ell}^{\ell}\psi_{m,\ell}~ Y_{\ell}^{m}(\phi, \lambda)
\label{eq:spherical-harmonics-decomposition}
\end{equation}
All fields $\omega,u,v$ and $\psi$ are $2\pi$-periodic in the zonal
($\lambda$) direction, so we can also define the Fourier coefficients
in the zonal direction, 
\begin{equation}
\psi_{m}(\phi)\equiv\frac{1}{2\pi}\int_{0}^{2\pi}\psi(\phi, \lambda)~ \mbox{e}^{-im\lambda}\,\mbox{d}\lambda=\sum_{\ell=|m|}^{\infty}\psi_{m,\ell}~ P_{\ell}^{m}(\sin \phi),\label{eq:Fourier-definition}
\end{equation}
with the associated Legendre polynomials $P_{\ell}^{m}(\sin \phi)$.  

In (\ref{eq:barotropic-topography-d}), $\kappa$ is a linear friction
coefficient, also known as Ekman drag or Rayleigh friction, that models
the dissipation of energy at the large scales of the flow \cite{vallis_atmospheric_2006}.
Hyper-viscosity $\nu_{n}\left(-\Delta\right)^{n}$ accounts for the
dissipation of enstrophy at small scales and is used mainly for numerical
reasons.  Most of the dynamical
quantities are independent of the value of $\nu_{n}$, for small enough
$\nu_{n}$. $\eta$ is a Gaussian noise with zero mean and correlations
$\mathbb{E}\left[\eta\left(\lambda_{1},\phi_{1},t_{1}\right)\eta\left(\lambda_{2},\phi_{2},t_{2}\right)\right]=C\left(\lambda_{1}-\lambda_{2},\phi_{1},\phi_{2}\right)\delta\left(t_{1}-t_{2}\right)$,
where $C$ is a positive-definite function and $\mathbb{E}$ is the expectation over realizations of the noise $\eta$. $C$ is assumed to be
normalized such that $\sigma$ is the average injection of energy
per unit of time and per unit of mass by the stochastic force $\sqrt{\sigma}\eta$. There is no symmetry reason to enforce homogeneous forcing over a rotating sphere, which only has axial symmetry.  Thus it is natural to consider forcing that varies with latitude.  The barotropic equation is sometimes used to describe the vertically-averaged atmospheric dynamics. The stochastic forces  model the driving influence of the baroclinic instability on the barotropic flow.  Baroclinic instabilities are typically strongest at mid-latitude.

%%%%%%%%%%%%%%%%%%%%%%%%%%%%%%%%
\subsection{Time scale separation between large scale and small scale dynamics}

%%%%%%%%%%%%%%%%%%%%%%%%%%%%%%%%
\subsubsection{Energy balance and non--dimensional equations\label{sec:LD-energy-balance-sphere}}

The inertial barotropic model (eq. (\ref{eq:barotropic-topography-d})
with $\kappa=\nu_{n}=\sigma=0$) conserves the energy $\mathcal{E}\left[\omega\right]=-\frac{1}{2}\int\omega\psi\,\mbox{d}{\bf r}$
(we denote by $\mbox{d}{\bf r}=a^{2}\cos\phi\,\mbox{d}\phi\mbox{d}\lambda$),
the moments of potential vorticity $\mathcal{C}_{m}\left[\omega\right]=\int(\omega+f)^{m}\,\mbox{d}{\bf r}$
with the Coriolis parameter $f(\phi)=2\Omega\sin\phi$, and the angular
momentum $L[\omega]=\int\omega\cos\phi\,\mbox{d}{\bf r}$.
The average energy balance for the dissipated and stochastically forced
barotropic equation is obtained applying the Ito formula \cite{Gardiner_1994_Book_Stochastic} to \eqref{eq:barotropic-topography-d}. It reads 
\begin{equation}
\frac{dE}{dt}=-2\kappa E-2\nu_{n}Z_{n}+\sigma,\label{eq:energy-balance-visc}
\end{equation}
where $E=\mathbb{E}\left[\mathcal{E}\left[\omega\right]\right]$ is
the total average energy and $Z_{n}=\mathbb{E}\left[-\frac{1}{2}\int\psi(-\Delta)^{n}\omega\,\mbox{d}{\bf r}\right]$.
The term $-2\nu_{n}Z_{n}$ in \eqref{eq:energy-balance-visc} represents
the dissipation of energy at the small scales of the flow. In the
regime we are interested in, most of the energy is concentrated in
the large-scale zonal jet, so the main mechanism of energy dissipation
is the linear friction (first term in the right-hand side of \eqref{eq:energy-balance-visc}).
In this turbulent regime, energy dissipated by hyper-viscosity
can be neglected. Then, in a statistically stationary state, $E_{stat}\simeq\frac{\sigma}{2\kappa}$,
expressing the balance between stochastic forces and linear friction
in \eqref{eq:barotropic-topography-d}.

The estimated total energy yields a typical jet velocity of
$U\sim\sqrt{\frac{\sigma}{2\kappa}}$. The order of magnitude of the
time scale of advection and stirring of turbulent eddies by this jet
is $\tau_{eddy}\sim\frac{a}{U}$. We perform a non-dimensionalization
of the stochastic barotropic equation \eqref{eq:barotropic-topography-d}
using $\tau_{eddy}$ as unit time and $a$ as unit length. The non-dimensionalization may be carried out by 
setting $a=1$ and using the non-dimensionalized variables $t'=t/\tau_{eddy}$,
$\omega'=\omega\tau_{eddy}$, $\psi'=\psi\tau_{eddy}$, $\Omega'=\Omega\tau_{eddy}$,
\begin{equation}
\alpha=\kappa\tau_{eddy}=\sqrt{\frac{2\kappa^{3}}{\sigma}},\label{eq:alpha}
\end{equation}
$\nu_{n}'=\nu_{n}\tau_{eddy}$, $\sigma'=\sigma\tau_{eddy}^{3}=2\alpha$,
and a rescaled force $\eta'=\eta\sqrt{\tau_{eddy}}$ such that $\mathbb{E}\left[\eta'\left(\lambda_{1},\phi_{1},t'_{1}\right)\eta'\left(\lambda_{2},\phi_{2},t'_{2}\right)\right]=C\left(\lambda_{1}-\lambda_{2},\phi_{1},\phi_{2}\right)\delta\left(t'_{1}-t'_{2}\right)$.
In these new units, and dropping the primes for simplicity,
the stochastic barotropic equation \eqref{eq:barotropic-topography-d}
reads
\begin{equation}
\frac{\partial\omega}{\partial t}+J(\psi,\omega)+2\Omega\frac{\partial\psi}{\partial\lambda}=-\alpha\omega-\nu_{n}\left(-\Delta\right)^{n}\omega+\sqrt{2\alpha}\eta.\label{eq:barotropic}
\end{equation}
In \eqref{eq:barotropic}, $\alpha$ is an inverse Reynolds' number
based on the linear friction and $\nu_{n}$ is an inverse Reynolds'
number based on hyper-viscosity. The turbulent regime mentioned previously
corresponds to $\nu_{n}\ll\alpha\ll1$. In such regime and in the
units of \eqref{eq:barotropic}, the total average energy in a statistically
stationary state is $E_{stat}=1$.

We are interested in the dynamics of zonal jets in the regime of small
forces and dissipation, defined as $\alpha\ll1$. 
In the next section we show that the dynamics corresponds to a regime in which the zonal jet evolves much more slowly
than the surrounding turbulent eddies.

%%%%%%%%%%%%%%%%%%%%%%%%%%%%%%%%
\subsubsection{Decomposition into zonal and non--zonal components}

In order to decompose \eqref{eq:barotropic} into a zonally averaged
flow and perturbations around it, we define the zonal projection of
a field 
\[
\left\langle \psi\right\rangle (\phi)\equiv\psi_{0}(\phi)=\frac{1}{2\pi}\int_{0}^{2\pi}\psi(\lambda,\phi)\,\mbox{d}\lambda.
\]
The zonal jet velocity profile is  defined by $U(\phi)\equiv\left\langle u\right\rangle (\phi)$.
In most situations of interest, the stochastic force in \eqref{eq:barotropic}
does not act direcly on the zonal flow: $\left\langle \eta\right\rangle =0$.
Then the perturbations of the zonal jet is proportional to the
amplitude of the stochastic force in \eqref{eq:barotropic}. We thus
decompose the velocity field as ${\bf v}=U{\bf e}_{x}+\sqrt{\alpha}\delta{\bf v}$
and the relative vorticity field as $\omega=\omega_{z}+\sqrt{\alpha}\delta\omega$
with $\omega_{z}\equiv\left\langle \omega\right\rangle $, where
$\alpha$ is the non-dimensional parameter defined in \eqref{eq:alpha}.  We call the perturbation velocity $\delta{\bf v}$ and vorticity 
$\delta\omega$ the eddy velocity and eddy vorticity, respectively.\\

With the decomposition of the vorticity field, the barotropic equation
\eqref{eq:barotropic} reads
\begin{equation}
\left\lbrace \begin{aligned} & \frac{\partial\omega_{z}}{\partial t}=\alpha R-\alpha\omega_{z}-\nu_{n}\left(-\Delta\right)^{n}\omega_{z}\\
 & \frac{\partial\delta\omega}{\partial t}=-L_{U}\left[\delta\omega\right]-\sqrt{\alpha}NL\left[\delta\omega\right]+\sqrt{2}\eta,
\end{aligned}
\right.\label{eq:barotropic-decomposed}
\end{equation}
with 
\begin{equation}
R(\phi)\equiv-\left\langle J\left(\delta\psi,\delta\omega\right)\right\rangle
\label{eq:R}
\end{equation}
the zonally averaged advection term,
where the linear operator $L_{U}$ reads
\begin{equation}
L_{U}\left[\delta\omega\right]=\frac{1}{\cos\phi}\left(U(\phi)\partial_{\lambda}\delta\omega+\gamma(\phi)\partial_{\lambda}\delta\psi\right)+\alpha\delta\omega+\nu_{n}\left(-\Delta\right)^{n}\delta\omega,\label{eq:LD-linear-operator}
\end{equation}
with $\gamma\left(\phi\right)=\partial_{\phi}\omega_{z}(\phi)+2\Omega\cos\phi$,
and where
\[
NL\left[\delta\omega\right]=J(\delta\psi,\delta\omega)-\left\langle J(\delta\psi,\delta\omega)\right\rangle 
\]
is the non-linear eddy-eddy interaction term.\\

Using $\omega_{z}\left(\phi\right)=-\frac{1}{\cos\phi}\partial_{\phi}\left(U\left(\phi\right)\cos\phi\right)$
and the first equation of (\ref{eq:barotropic-quasi-linear}), we
get the evolution equation for the zonal flow velocity $U\left(\phi\right)$
\begin{equation}
\frac{\partial U}{\partial t}=\alpha f-\alpha U-\nu_{n}\left(-\Delta\right)^{n}U\,,\label{eq:zonal-vorticity-ell}
\end{equation}
where $f\left(\phi\right)$ is such that $R\left(\phi\right)=-\frac{1}{\cos\phi}\partial_{\phi}\left(f\left(\phi\right)\cos\phi\right)$. $f$ is minus the divergence of the Reynolds' stress.

%%%%%%%%%%%%%%%%%%%%%%%%%%%%%%%%
\subsubsection{Quasi-linear and linear dynamics}

In this section we discuss the quasilinear approximation to the barotropic equation and the associated linear dynamics.

In the limit of small forces and dissipation $\alpha\ll1$, the perturbation
flow is expected to be of small amplitude. Then the non-linear term
$NL[\delta\omega]$ in (\ref{eq:barotropic-decomposed}) is negligible
compared to the linear term $L_{U}\left[\delta\omega\right]$. Neglecting
these non-linear eddy-eddy interaction terms, we obtain the so-called
quasi-linear approximation of the barotropic equation \cite{Srinivasan-Young-2011-JAS},
\begin{equation}
\left\lbrace \begin{aligned} & \frac{\partial\omega_{z}}{\partial t}=\alpha R-\alpha\omega_{z}-\nu_{n}\left(-\Delta\right)^{n}\omega_{z}\\
 & \frac{\partial\delta\omega}{\partial t}=-L_{U}\left[\delta\omega\right]+\sqrt{2}\eta.
\end{aligned}
\right.\label{eq:barotropic-quasi-linear}
\end{equation}
The approximation leading to the quasi-linear dynamics (\ref{eq:barotropic-quasi-linear})
amounts at suppressing some of the triad interactions. Nonetheless,
the inertial quasi-linear dynamics has the same quadratic invariants
as the initial barotropic equations. The average energy balance for
the quasi-linear barotropic dynamics (\ref{eq:barotropic-quasi-linear})
is thus the same as the one for the full barotropic dynamics \eqref{eq:barotropic-decomposed}.\\

For many flows of interest, for example Jovian
jets, the turbulent eddies $\delta\omega$ evolve much faster than
the zonal jet velocity profile $U$ \cite{Porco2003}. In (\ref{eq:barotropic-decomposed})
and (\ref{eq:barotropic-quasi-linear}), the natural time scale of
evolution of the zonal jet is of order $1/\alpha$, while the typical
time scale of evolution of the perturbation vorticity $\delta\omega$
is of order $1$. In the regime $\alpha\ll1$, we thus expect to observe
a separation of time scales between the evolution of $\omega_{z}$
and $\delta\omega$, consistent with the definition of $\alpha$
as the ratio of the inertial time scale $\tau_{eddy}$ and of the
dissipative time scale $1/\kappa$, see~(\ref{eq:alpha}).

In the regime $\alpha\ll1$, it is natural to consider the linear dynamics of $\delta\omega$ with $U$ held fixed,
\begin{equation}
\frac{\partial\delta\omega}{\partial t}=-L_{U}\left[\delta\omega\right]+\sqrt{2}\eta\,.\label{eq:frozen-QL}
\end{equation}
The relevance of \eqref{eq:frozen-QL} as an effective description
of turbulent eddy dynamics is further discussed later.
In particular, we show in section \ref{sub:Empirical-validation-of}
that the correlation time of Reynolds' stresses resulting from the
linear dynamics \eqref{eq:frozen-QL} ---the most relevant time scale
related to the dynamics of eddies and their action on the evolution of the zonal jet---
is of the order or smaller than $\tau_{eddy}$, holding even as $\alpha$ decreases. It means that the time scale separation
hypothesis that leads us to consider the linear dynamics \eqref{eq:frozen-QL}
is self-consistent in the limit of weak forces and dissipation $\alpha\ll1$.

\subsubsection{Reynolds averaging for the vorticity equation }

In the introduction we discussed Reynolds averaging and Reynolds stresses for the simplest possible case: a two dimensional flow that does not break the symmetry along the direction $\mathbf{e}_{x}$. We now adapt the discussion to two dimensional flows on a sphere.  As it is much more convenient to work directly with the vorticity equation, we discuss Reynolds averaging for the vorticity equation only. 

Our aim is to write the counterpart of Eq. \eqref{eq:Slow_Stochastic_Dynamics} and \eqref{eq:Time_Averaged_Reynold_Stress}, for the vorticity equation. In the cases when there is a time scale separation between the evolution of the slow zonal and the fast non zonal part of the flow, averaging either Eq. \eqref{eq:barotropic-decomposed} or Eq. \eqref{eq:barotropic-quasi-linear} leads to an effective equation for the low frequency evolution of the zonal vorticity
\begin{equation}
\frac{\partial\omega_{z}}{\partial t}=\alpha \mathbb{E}\left(R\right)-\alpha\omega_{z}-\nu_{n}\left(-\Delta\right)^{n}\omega_{z} + \xi_{\omega_z},
\label{R_gaussian_evolution}
\end{equation}
where $\mathbb{E}\left(R\right)$ is the average of the vorticity flux $R$ \eqref{eq:R}, and the white in time Gaussian noise $\xi_{\omega}$ describes the typical fluctuations. We consider time averages of the vorticity flux
\begin{equation}
r=\frac{1}{T}\int\text{dt}\,R(u).\label{eq:Reynold_Stress_fluctuations_vorticity}
\end{equation}
The average of $r$ is the term $\mathbb{E}\left(R\right)$ appearing in the Reynolds averaged equation \eqref{R_gaussian_evolution}. We call this term the vorticity Reynolds stress; however it does not have the same physical dimension as the usual stress. When the time average is over a time window of duration $T$ which
is assumed to be short compared to the time scale for the evolution of $U$, but large compared with the evolution of the turbulent fluctuations:
$\tau_{e}\ll T\ll\tau_{U}$,  we call the fluctuations of (\ref{eq:Reynold_Stress_fluctuations_vorticity}) the vorticity Reynolds stress fluctuations (the fluctuation of the time averaged vorticity fluxes, over finite but long times $T$).  In the asymptotic regime $\tau_{e}\ll T$, the probability distribution
function of $r$ takes the simple large deviation form $P(r,T)\underset{T\rightarrow\infty}{\asymp}\exp\left(-TI[r])\right)$.
The variance of $\xi_{\omega}$ is given by a Kubo formula, and is simply related to the second variations of $I$. 

We note that there exists a simple relation between the Reynolds stress large deviations rate function $I_v$, that describes the averages of the actual momentum fluxes that appear in the velocity equation, and the vorticity Reynolds stress large deviation rate function $I$. In the following we study the vorticity Reynolds stress only. For simplicity, as there is no ambiguity, we call these quantities Reynolds stresses and Reynolds stress large deviation rate functions, omitting the word vorticity.

%%%%%%%%%%%%%%%%%%%%%%%%%%%%%%%%
\subsection{Numerical implementation\label{sub:Numerical-implementation}}

Direct numerical simulations (DNS) of the barotropic equation (\ref{eq:barotropic-decomposed}),
the quasi-linear barotropic equation (\ref{eq:barotropic-quasi-linear})
and the linear equation (\ref{eq:frozen-QL}) are performed using
a purely spectral code with a fourth-order-accurate Runge-Kutta algorithm
and an adaptive time step\footnote{A program that implements spectral DNS for the non-linear
and quasi-linear equations, solves the non-linear Riccati equation, and includes graphical tools
to visualize statistics, is freely available.  The application
``GCM'' is available for OS X 10.9 and higher on
the Apple Mac App Store at URL http://appstore.com/mac/gcm}. The spectral cutoffs defined by $\ell\leq L$, $\left|m\right|\leq\min\left\{ \ell,M\right\} $
in the spherical harmonics decomposition of the fields are taken to
be $L=80$ and $M=20$. In all the simulations, the rotation rate
of the sphere is $\Omega=3.7$ in the units defined previously.

The stochastic noise is implemented using the method proposed in Ref. \onlinecite{lilly1969numerical},
with a non-zero renewal time scale $\tau_{r}$ larger than the time
step of integration. For $\tau_{r}$ much smaller than the typical
eddy turnover time scale, the noise can be considered as white in
time.

Whenever one considers the linear dynamics (\ref{eq:frozen-QL}), modes with different values of $m$ decouple, thanks to the zonal symmetry. Then the statistics of the contribution of the Reynolds stress coming from different values of $m$ are independent. The statistics for the total Reynolds stress can be computed from the statistics of the contribution of each independent value of $m$. It is  natural and simpler to study the contribution from each different value of $m$ independently. For this reason we consider in this study a force that acts on one mode only. However, as explained in the previous section the validity of the quasilinear approximation is favored by the use of a broad band spectrum forcing, or a forcing acting on a large number of small scale modes, or both.  Forcing only one mode is the most unfavorable case from the point of view of the accuracy of the quasilinear approximation.  Larger time scale separation may be required in this case to ensure the accuracy of the quasilinear approximation.  However whenever the quasilinear approximation is accurate, the statistics of the Reynolds stress arising from the forced mode are accurately described by the methods reported here.

\begin{figure}\begin{center}
\includegraphics[scale=0.5]{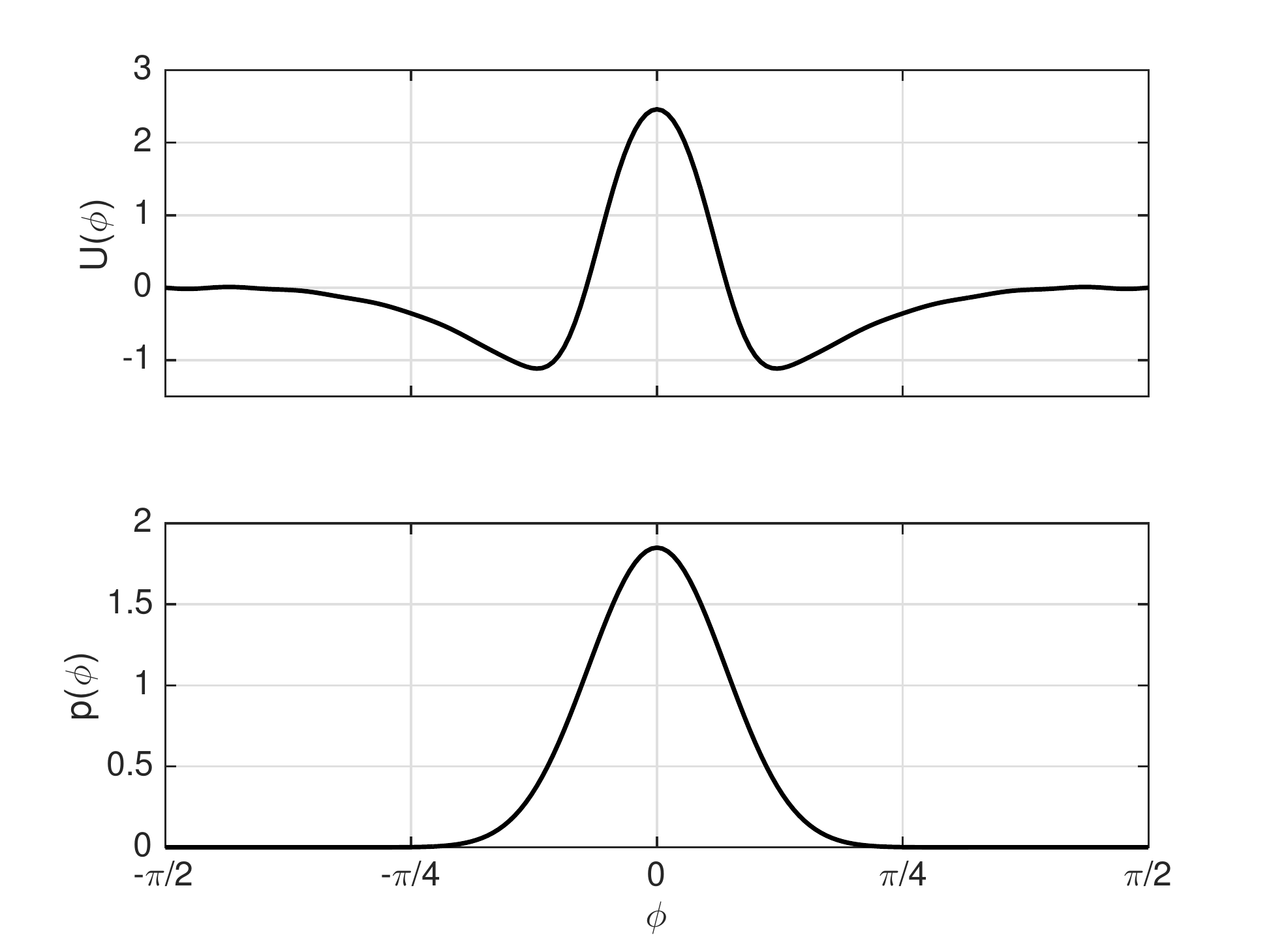}

\protect\caption{\label{fig:U}Top pannel: the zonal flow velocity profile $U\left(\phi\right)$
used in numerical simulations of the linearized barotropic equation
\eqref{eq:frozen-QL}. Bottom panel: zonally averaged energy injection rate by the stochastic force $\eta$ in \eqref{eq:barotropic}, \eqref{eq:barotropic-quasi-linear} and \eqref{eq:frozen-QL}.}

\end{center}\end{figure}

The forcing only acts on the mode $\left|m\right|=10$, $\ell=10$,
which is concentrated around the equator (see figure \ref{fig:U}).  With such a forcing spectrum
and setting $\alpha=0.073$, the integration of the quasi-linear barotropic
equation (\ref{eq:barotropic-quasi-linear}) leads to a stationary
state characterized by a strong zonal jet with velocity $U\left(\phi\right)$, represented in Figure \ref{fig:U}.  We spectrally truncate the 
jet to its first 25 spherical harmonics to fix the mean flow in the simulation
of the linear barotropic equation (\ref{eq:frozen-QL}).
We use hyper-viscosity of order 4 with coefficient $\nu_{4}$ such that the damping rate of the smallest mode is 4. To assess that hyper-viscosity
is negligible in the large scale statistics, simulations of the linear
equation with $\nu_{4}=4$ and $\nu_{4}=2$ are compared in sections \ref{sec:Equal-time-statistics-of}, \ref{sub:Gaussian-approximation-of} and \ref{sub:Application-of-the}.

%%%%%%%%%%%%%%%%%%%%%%%%%%%%%%%%
%%%%%%%%%%%%%%%%%%%%%%%%%%%%%%%%
%%%%%%%%%%%%%%%%%%%%%%%%%%%%%%%%
\section{Equal-time statistics of vorticity fluxes\label{sec:Equal-time-statistics-of}}

\begin{figure}
\includegraphics[scale=0.5]{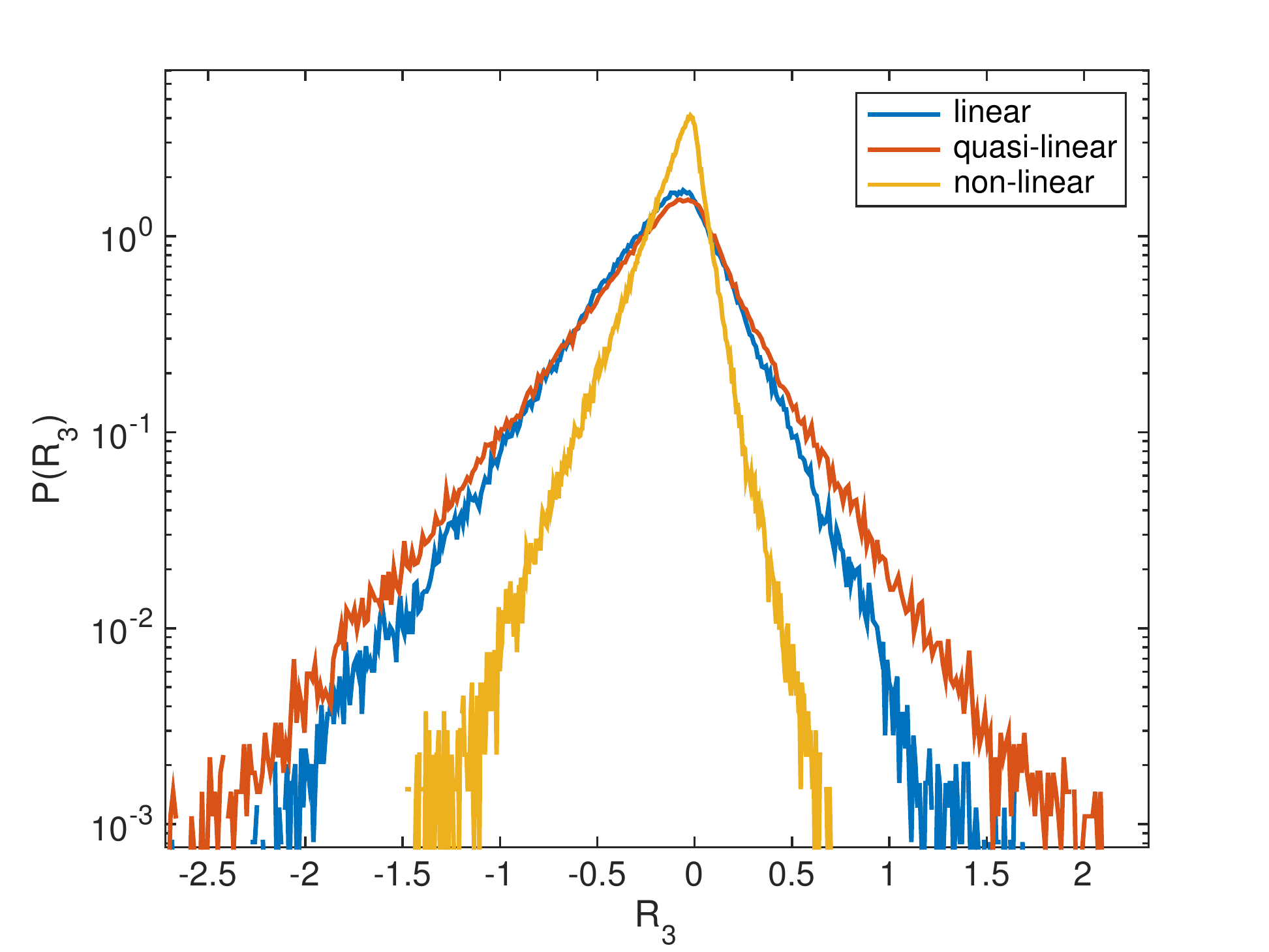}

\protect\caption{\label{fig:histograms}Probability density functions of $R_{3}$, the third component
in the spherical harmonics decomposition of the zonally averaged advection term (vorticity flux) $R(\phi)$, from
direct numerical simulations of the linear barotropic equation \eqref{eq:frozen-QL}
(blue), the quasi-linear barotropic equation \eqref{eq:barotropic-quasi-linear}
(orange), and the non-linear barotropic equation \eqref{eq:barotropic-decomposed}
(yellow). Exponential tails are observed in all of the different
cases. The common parameters are $\alpha=0.073$, $\Omega=3.7$, total
integration time $5,450$, and the forcing is concentrated in wavenumbers
$\left|m\right|=10$, $\ell=10$.}
\end{figure}

\begin{figure}
\includegraphics[scale=0.4]{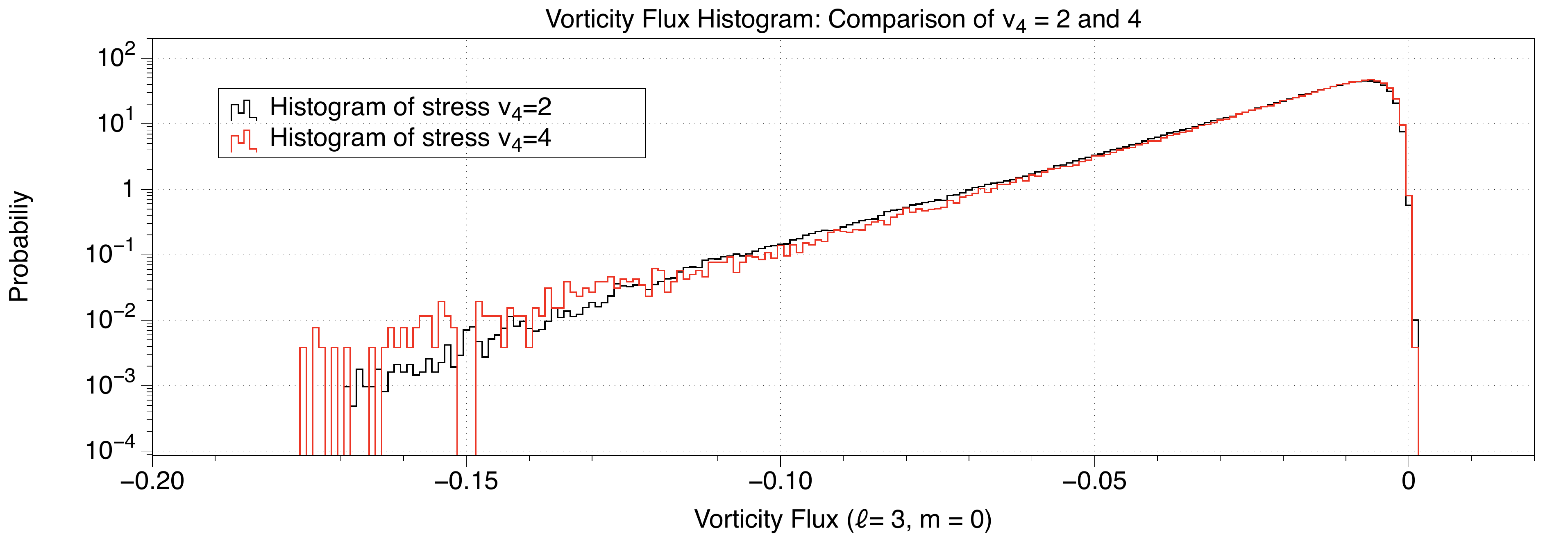}

\protect\caption{\label{fig:histogram-viscosity}Probability density functions of $R_{3}$, the third
component in the spherical harmonics decomposition of the zonally averaged advection term (vorticity flux) $R$, from direct numerical simulations of the quasi-linear barotropic
equation \eqref{eq:barotropic-quasi-linear} with hyper-viscosity
such that the smallest scale has a hyperviscous damping rate of $4$
(red curve) and $2$ (black curve).  The two probability density functions are nearly identical, showing
that hyper-viscosity can be considered to be negligible as far as the zonal jet
statistics are concerned.}
\end{figure}

The aim of this section is to illustrate that fluctuations of equal-time vorticity flux $R$ \eqref{eq:R} may be strongly non Gaussian. We prove that vorticity flux fluctuations have exponential tails
with a distribution close to that of Gaussian product statistics \cite{Grooms:2016kw}. While equal-time fluctuations of the vorticity flux are important for high frequency jet variability, Reynolds stresses (time average of the vorticity fluxes) are more important for the long term evolution of the jet. Beginning in section IV, we study Reynolds stresses, and their large deviations.

The evolution of the mean flow $\omega_{z}(\phi,t)$ is given by the
advection term $R(\phi,t)=-\left\langle J\left(\delta\psi,\delta\omega\right)\right\rangle $,
through (\ref{eq:barotropic-decomposed}) or (\ref{eq:barotropic-quasi-linear}).
In most previous statistical approaches to zonal jet dynamics, only
the averaged advection term, the Reynolds stress, was considered.
This is for instance the case in S3T \cite{bakas2015s3t} and CE2
\cite{marston65conover,Srinivasan-Young-2011-JAS,tobias2013direct}
approaches. 
Such restriction gives a good approximation of the relaxation
of zonal jets towards the attractors of the dynamics, that is expected
to be quantitatively accurate in the inertial limit $\alpha\to0$
\cite{Bouchet_Nardini_Tangarife_2013_Kinetic}. However, replacing
the advection term $R$ by its average does not describe fluctuations
of the vorticity fluxes, that may lead to fluctuations of zonal jets.
Understanding the statistics of vorticity fluxes beyond their average
value is thus a very interesting perspective. In this section, we
study the whole distribution function of vorticity fluxes, as computed
from direct numerical simulations.\\

The zonally averaged advection term is a function of latitude $\phi$ and can be decomposed with spherical harmonics
 according to \eqref{eq:spherical-harmonics-decomposition}.
We denote by $R_{\ell}(t)\equiv R_{0,\ell}(t)$ the $\ell$-th component
in the spherical harmonics decomposition of $R(\phi,t)$. All $R_{l}$ for odd values of $l$ larger than one have non-zero amplitudes (the amplitude of the $l=1$ mode is zero because total angular momentum about the polar axis remains zero). In the following, for simplicity,  we focus our analysis on $R_{3}$ only, that has the largest contribution. The probability density functions of $R_{3}$, computed either
from direct numerical simulations of the barotropic equation \eqref{eq:barotropic-decomposed},
or the quasi-linear barotropic equation \eqref{eq:barotropic-quasi-linear}
or the linear equation \eqref{eq:frozen-QL}, with the forcing spectrum
specified in section \ref{sub:Numerical-implementation} and with
$\alpha=0.073$, are shown in Figure
\ref{fig:histograms}. Figure \ref{fig:histogram-viscosity} shows that the probability distribution of $R_3$ is not affected by the choice of small scale dissipation.

In the linear dynamics \eqref{eq:frozen-QL}, the eddy vorticity evolves
according to the linearized barotropic equation close to the fixed
base flow $U(\phi)$ shown in Figure \ref{fig:U}. In the quasi-linear
dynamics \eqref{eq:barotropic-quasi-linear}, the zonal mean flow
has the same average velocity profile $U(\phi)$, but this zonal flow
is allowed to fluctuate. This difference in the dynamics of the zonal
flow between linear and quasi-linear equations explains the slight
difference observed in the corresponding advection term histograms
(respectively blue curve and orange curve in Figure \ref{fig:histograms}),
namely, the probability density function is more spread (the vorticity fluxes fluctuate
more) in the quasi-linear dynamics than in the linear dynamics.

In contrast, the probability density function of $R_{3}$ computed from the non-linear
integration (yellow curve in Figure \ref{fig:histograms}) is very
different from the other ones for two reasons: the average zonal flow
is different from the fixed zonal flow used in the linear dynamics,
and the dynamics of $\delta\omega$ is also different from the quasi-linear
dynamics because of the non-linear eddy-eddy interaction terms in
\eqref{eq:barotropic-decomposed} (this is expected, as forcing a single mode is the most unfavorable case from the point of view of the validity of the quasilinear approximation, as explained in section \ref{sec:Barotropic-equation-and}).\\

In all three cases, the probability distribution functions in Figure \ref{fig:histograms} show large fluctuations and heavy tails. For instance it is clear that typical fluctuations of the vorticity flux have much larger amplitude than the value of their average (the variance is much larger than the average). While essential for understanding the high frequency and small variability of the jets, on the slow time scale, the jet evolution is described by time averaged vorticity fluxes (the Reynolds stress).  

In all of the simulations, the distribution of the vorticity flux
shows exponential tails. This can be easily understood for the case
of the linear equation \eqref{eq:frozen-QL}. Indeed, in this case
the statistics of the eddy vorticity are exactly Gaussian ($\delta\omega$
is an Ornstein-Uhlenbeck process \cite{Gardiner_1994_Book_Stochastic}).
Then, the statistics of $R(\phi)$ can be calculated explicitly, as
we explain now.

Using \eqref{eq:Fourier-definition} we can write the vorticity flux
as
\begin{equation}
R(\phi)=-\frac{1}{\cos\phi}\sum_{m}im\left(\psi_{m}\cdot\partial_{\phi}\omega_{-m}+\partial_{\phi}\psi_{m}\cdot\omega_{-m}\right),\label{eq:Reynolds-stress-m}
\end{equation}
where $\omega_{m}(\phi)$ is the $m$-th Fourier coefficient of $\delta\omega$,
and $\psi_{m}(\phi)$ is the associated stream function. The Ornstein-Uhlenbeck
process $\omega_{m}\left(\phi\right)$ is a Gaussian random variable
at each latitude $\phi$. The sum of Gaussian random variables is
a Gaussian random variable, so $\psi_{m}(\phi)$, $\partial_{\phi}\psi_{m}(\phi)$
and $\partial_{\phi}\omega_{m}(\phi)$ are also Gaussian random variables
at each latitude $\phi$. All these Gaussian random variables have
zero mean, and in general they are correlated in a non-trivial way.

The vorticity flux \eqref{eq:Reynolds-stress-m} is thus of the form
$R=\xi_{1}\xi_{2}+\ldots+\xi_{M-1}\xi_{M}$ where $\xi_{1},\ldots,\xi_{M}$
are $M$ real-valued\footnote{We can restrict ourselves to real $\xi_{m}$ decomposing $\omega_{m}$
and $\psi_{m}$ into real and imaginary parts.} correlated Gaussian variables with zero mean. We denote by $\xi$
the column vector with components $\xi_{1},\ldots,\xi_{M}$. By definition,
the probability distribution function of $\xi$ is 
\[
P_{\xi}\left(\xi\right)=\frac{1}{Z}\exp\left(-\frac{1}{2}\xi^{T}G^{-1}\xi\right),
\]
where $\xi^{T}$ denotes the transpose vector of $\xi$, $G$ is the
covariance matrix of $\xi$, and $Z$ is a normalisation constant.
The probability density function of $R$, denoted $P_{R}$, is given
by 
\begin{align*}
P_{R}(R) & =\int\mbox{d}\xi\,P_{\xi}\left(\xi\right)\delta\left(R-\xi_{1}\xi_{2}-\ldots-\xi_{M-1}\xi_{M}\right)\\
 & =\int\mbox{d}\xi_{2}\ldots\mbox{d}\xi_{m}\,\frac{1}{\left|\xi_{2}\right|}\,P_{\xi}\left(\frac{R-\xi_{3}\xi_{4}-\ldots-\xi_{M-1}\xi_{M}}{\xi_{2}},\xi_{2},\ldots\xi_{M}\right).
\end{align*}
Using the change of variable $\zeta_{m}=\xi_{m}/\sqrt{\left|R\right|}$
for $m=2,\ldots,M$, the first argument of $P_{\xi}$ becomes $\sqrt{\left|R\right|}\frac{\frac{R}{\left|R\right|}-\zeta_{3}\zeta_{4}-\ldots-\zeta_{M-1}\zeta_{M}}{\zeta_{2}}$,
so we obtain: 
\[
P_{R}(R)=\frac{1}{Z}\int\mbox{d}\zeta_{2}\ldots\mbox{d}\zeta_{M}\,\frac{\left|R\right|^{\frac{M-2}{2}}}{\left|\zeta_{2}\right|}\,\exp\left(-\left|R\right|Q_{\pm}\left(\zeta_{2},\ldots,\zeta_{M}\right)\right),
\]
where $Q_{\pm}$ is a function of $\left(\zeta_{2},\ldots,\zeta_{M}\right)$,
that depends only on the sign of $R$, according to $R=\pm\left|R\right|$.
The tails of the distribution $P_{R}$ correspond to the limits $R\to\pm\infty$.
In both limits, $\left|R\right|\to\infty$ so we can perform a saddle-point
approximation in the above integral, and get 
\begin{equation}
\ln\left(P_{R}(R)\right)\underset{R\to\pm\infty}{\sim}-\left|R\right|\mu_{\pm},\label{eq:PDF-exponential-tails}
\end{equation}
where the rates of decay are defined by 
\begin{equation}
\mu_{\pm}=\min_{\zeta_{2},\ldots,\zeta_{M}}\left\{ Q_{\pm}\left(\zeta_{2},\ldots,\zeta_{M}\right)\right\} .\label{eq:exponential-rate}
\end{equation}
The exponential tails of the distribution $P_{R}$
are direct consequences of the fact that the eddy vorticity $\delta\omega$
evolving according to the linear equation \eqref{eq:frozen-QL} is
a Gaussian process and of the fact that $R$ is quadratic in $\delta\omega$.
This simple argument explains the exponential tails observed in probability density functions
of the zonally averaged advection term in simulations of the linear dynamics \eqref{eq:frozen-QL}
(blue curve in Figure \ref{fig:histograms}), where the vorticity
field is exactly an Ornstein-Uhlenbeck process.\\

In the quasi-linear and non-linear dynamics, the zonal flow and eddies
evolve at the same time scale. As a consequence, the dynamics of the eddy
vorticity is not linear, and its statistics are not Gaussian. However,
we observe that the probability density functions of eddy vorticity are
nearly Gaussian (skewness -0.0147 and kurtosis 3.8079 in the quasi-linear
case, skewness -0.0037 and kurtosis 3.3964 in the non-linear case,
compared to skewness 0.0172 and kurtosis 3.0028 in the linear case).
The previous argument can thus also be applied empirically to explain
the exponential tails observed in the curves corresponding to quasi-linear
and non-linear simulations in Figure \ref{fig:histograms}. \\

\begin{figure}
\includegraphics[scale=0.6]{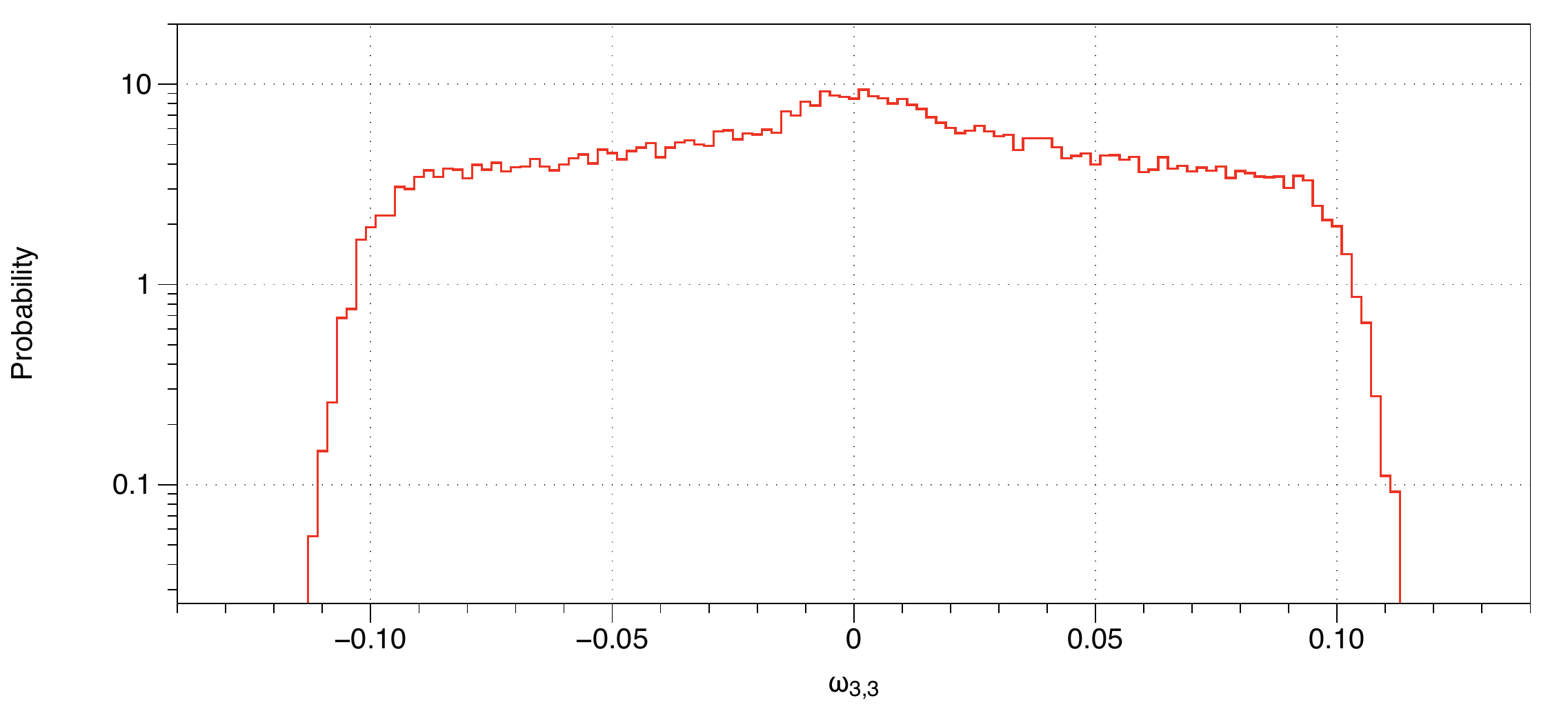}

\includegraphics[scale=0.6]{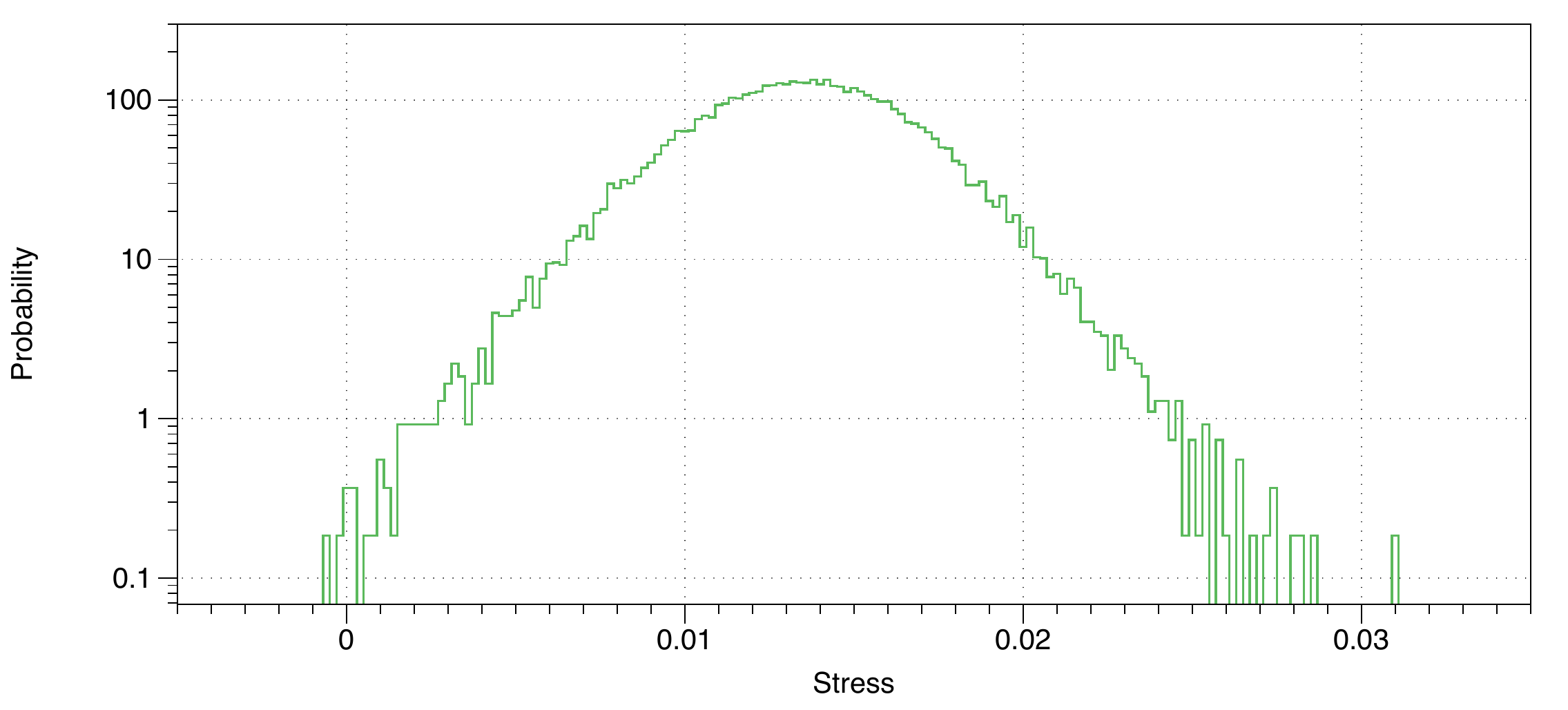}

\protect\caption{\label{fig:baroclinic}Probability density functions of the vorticity component $\omega_{3,3}$
(top panel) and zonally averaged advection term (vorticity flux) $R_{3}$ (bottom pannel) from a direct
numerical simulation of the deterministic 2-layer quasi-geostrophic
baroclinic equation. The eddy vorticity is clearly non-Gaussian, and
yet the advection term distribution has exponential tails as in the one-layer
cases (Figure \ref{fig:histograms}). This observation calls for a
more general study of vorticity flux statistics close to a zonal
jet.}
\end{figure}

The same analysis has been performed on direct numerical simulations
of the deterministic 2-layer quasi-geostrophic baroclinic model \cite{vallis_atmospheric_2006},
see Figure \ref{fig:baroclinic}. In this case, the eddy vorticity
statistics are highly non-Gaussian, while statistics of the vorticity flux
have exponential tails similar to those found in the one-layer case.
The observation indicates that the previous explicit calculation
might not be the most general explanation of the exponential distribution
of vorticity fluxes.

%%%%%%%%%%%%%%%%%%%%%%%%%%%%%%%%
%%%%%%%%%%%%%%%%%%%%%%%%%%%%%%%%
%%%%%%%%%%%%%%%%%%%%%%%%%%%%%%%%
\section{Averaging and large deviations in systems with time scale separation\label{sec:large_deviations}}

As explained in section \ref{sec:Barotropic-equation-and}, we are
interested in the regime where zonal jets evolve much slower than
the surrounding turbulent eddies. In this section, we present some
theoretical tools (stochastic averaging, large deviation principle)
that can be applied to study the effective dynamics and statistics
of slow dynamical variables coupled to fast stochastic processes.
Most of these tools are classical ones \cite{freidlin2012random,Gardiner_1994_Book_Stochastic,pavliotis2008multiscale},
except for the explicit results presented in section \ref{sub:Quasi-linear-systems-with}
\cite{BouchetTangarifeVandenEijnden2015}. Application of these general tools
to the quasi-linear barotropic model is considered in sections
\ref{sub:Gaussian-approximation-of} and \ref{sub:Application-of-the}.

Consider the stochastic dynamical system
\begin{equation}
\left\lbrace \begin{aligned} & \frac{dx}{dt}=\alpha f\left(x,y\right)\\
 & \frac{dy}{dt}=b\left(x,y\right)+\eta
\end{aligned}
\right.\label{eq:slow-general}
\end{equation}
 where $0<\alpha\ll1$, and where $\eta$ is a Gaussian random column
vector with zero mean and correlations $\mathbb{E}\left[\eta\left(t_{1}\right)\eta^{T}\left(t_{2}\right)\right]=C\delta\left(t_{1}-t_{2}\right)$
with the correlation matrix $C$. In the case we are interested in,
the random vector $y$ is actually the eddy vorticity field,
and $x$ is the zonal jet vorticity or velocity. For simplicity we use vector
notation $x=\left(x_{\ell}\right)_{1\leq\ell\leq L}$ 
in this section, the formal generalization to the field case is straightforward,
see sections \ref{sub:Gaussian-approximation-of} and \ref{sub:Application-of-the}.

In (\ref{eq:slow-general}), the variable $x$ typically evolves on
a time scale of order $1/\alpha$, while $y$ evolves on a time scale
of order 1. When there is a time scale separation between zonal jets
and eddies, defined by $\alpha\ll1$, the quasi-linear barotropic equation
(\ref{eq:barotropic-quasi-linear}) is a particular case of the system
(\ref{eq:slow-general}). Note however that in that case, dissipation
terms of order $\alpha$ are present in $b(x,y)$. The general results
presented in this section usually do not take into account such terms
\cite{freidlin2012random,Gardiner_1994_Book_Stochastic,pavliotis2008multiscale}.
As a consequence, in sections \ref{sub:Gaussian-approximation-of} and \ref{sub:Application-of-the} 
we make sure that our results do not depend on the dissipative
terms in the limit $\alpha\to0$.\\

The goal of stochastic averaging is to give an effective description
of the dynamics of $x$ over time scales of order $1/\alpha$, where
the effect of the fast process $y$ is averaged out. The effective
dynamics describes the attractors of $x$, the relaxation dynamics
towards these attractors and the small fluctuations around these attractors,
in the regime $\alpha\ll1$.  For quasi-geostrophic
zonal jets dynamics, stochastic averaging leads to a kinetic description
of zonal jets \cite{Bouchet_Nardini_Tangarife_2013_Kinetic}, related
to statistical closures of the dynamics (S3T \cite{bakas2015s3t}
and CE2 \cite{Srinivasan-Young-2011-JAS,tobias2011astrophysical,tobias2013direct}).

The effective dynamics
obtained through stochastic averaging or statistical closures is not
able to describe arbitrarily large fluctuations of the slow process
$x$. Such rare events are of major importance in the long-term dynamics
of $x$. For instance in the case where the system (\ref{eq:slow-general})
has several attractors, transitions between the attractors are governed
by large fluctuations of the system. The description of such transitions
(transition probability, typical transition path) cannot be done through
a stochastic averaging procedure.

Large deviation theory is a
natural framework to describe large fluctuations of $x$ in the regime
$\alpha\to0$. The large deviation principle \cite{freidlin2012random}
gives the asymptotic form of the probability density of paths $\left\{ x(t)\right\} _{0\leq t\leq T}$
when $\alpha\ll1$, with the effect of the fast process $y$ averaged
out.  Information about the typical
effective dynamics of $x$ as obtained through stochastic averaging is captured, 
but the principle allows us to go further to describe arbitrarily rare events.
In cases of multistability of $x$, the Large Deviation Principle
yields the asymptotic expression of the transition probability from
one attractor to another, the average relative residence time in each
attractor, and the typical transition path $\left\{ x(t)\right\} _{0\leq t\leq T}$
that links two attractors in a given time $T\gtrsim1/\alpha$, among
other relevant statistical quantities. Implementing the large deviation
principle in practice for systems like (\ref{eq:slow-general}) and for the quasilinear dynamics is
one of the goals of this work.\\

In the effective descriptions of $x$ provided by stochastic averaging
and the Large Deviation Principle, the dynamics of $y$ is approximated
by its stationary dynamics with $x$ held fixed, the so-called virtual
fast process. The mathematics is described in section \ref{sub:The-virtual-fast}.
The effective dynamics of $x$ over time scales $t\gg1$ provided
by stochastic averaging is  presented in section \ref{sub:Average-evolution-and}.
The Large Deviation Principle for (\ref{eq:slow-general}) is stated
in section \ref{sub:Large-time-large-deviations}, and in section
\ref{sub:Sampling-the-SCGF} we give a method to estimate the quantities
involved in the Large Deviation Principle from simulations of the
virtual fast process.

\subsection{The virtual fast process\label{sub:The-virtual-fast}}

In slow-fast systems like (\ref{eq:slow-general}), the time scale
separation implies that at leading order, the statistics of $y$ are
very close to the stationary statistics of the virtual fast process
$\tilde{y}(u)$ 
\begin{equation}
\frac{d\tilde{y}}{du}=b\left(x,\tilde{y}(u)\right)+\eta(u),\label{eq:virtual-fast-process}
\end{equation}
where $x$ is held fixed \cite{freidlin2012random,Gardiner_1994_Book_Stochastic}. The time scale separation hypothesis is relevant only when the fast process described by \eqref{eq:virtual-fast-process} is stable (for instance has an invariant measure and is ergodic). The stationary process \eqref{eq:virtual-fast-process} depends parametrically on $x$, and the expectation over the invariant measure of \eqref{eq:virtual-fast-process} is thus denoted $\mathbb{E}_x$. The statistics of $\tilde{y}$  change when $x$ evolves adiabatically on longer timescales.

For  quasilinear barotropic dynamics \eqref{eq:barotropic-quasi-linear},
the virtual fast process is the linearized barotropic equation close
to the fixed stable zonal flow $U$ \eqref{eq:frozen-QL} (the necessity for $U$ to be stable for the quasilinear hypothesis to be correct was emphasized in reference \cite{Bouchet_Nardini_Tangarife_2013_Kinetic}.)\\

The process (\ref{eq:virtual-fast-process}) is relevant only if a
time scale separation effectively exists between the evolutions of
$x$ and $y$. In practice, the time scale separation hypothesis in
(\ref{eq:slow-general}) can be considered to be self-consistent if
the typical time scale of evolution of the virtual fast process (\ref{eq:virtual-fast-process})
is of order one, while the slow variable evolves on a time scale of order $1/\alpha$. From the point of view of the interaction with the dynamics of $x$, the
most relevant time scales related to the evolution of $\tilde{y}(u)$
are the correlation times of processes $f_{\ell}\left(x,\tilde{y}(u)\right)$
and $f_{\ell'}\left(x,\tilde{y}(u)\right)$, defined as \cite{newman1999monte,papanicolaou1977introduction}
\begin{align}
\tau_{\ell,\ell'} & =\lim_{t\to\infty}\frac{1}{t}\int_{0}^{t}\int_{0}^{t}\frac{\mathbb{E}_{x}\left[\left[f_{\ell}\left(x,\tilde{y}\left(u_{1}\right)\right)f_{\ell'}\left(x,\tilde{y}\left(u_{2}\right)\right)\right]\right]}{2\mathbb{E}_{x}\left[\left[f_{\ell}\left(x,\tilde{y}\right)f_{\ell'}\left(x,\tilde{y}\right)\right]\right]}\,\mbox{d}u_{1}\mbox{d}u_{2}\label{eq:def-autocorrelation-time}
\end{align}
where $\mathbb{E}_{x}\left[\left[X_{1}\left(u_{1}\right)X_{2}\left(u_{2}\right)\right]\right]\equiv\mathbb{E}_{x}\left[X_{1}\left(u_{1}\right)X_{2}\left(u_{2}\right)\right]-\mathbb{E}_{x}\left[X_{1}\left(u_{1}\right)\right]\mathbb{E}_{x}\left[X_{2}\left(u_{2}\right)\right]$
is the covariance of $X_{1}$ at time $u_{1}$ and $X_{2}$ at time
$u_{2}$. If $\ell=\ell'$, $\tau_{\ell,\ell}$ is called the auto-correlation
time of the process $f_{\ell}\left(x,\tilde{y}(u)\right)$. In all
these expressions, $x$ is fixed and $\mathbb{E}_{x}$ is the average
over realizations of the fast process (\ref{eq:virtual-fast-process})
in its statistically stationary state. The correlation times $\left\{ \tau_{\ell,\ell'}\right\} $
give an estimate of the time scales of evolution of the terms that
force the slow process $x$ in (\ref{eq:slow-general}).\\

%In the case of the linearized barotropic dynamics \eqref{eq:frozen-QL}, dissipative terms of order $\alpha$ are still present in the dynamics of the virtual fast process. Then, we have to make sure that $\tau_{\ell,\ell'}$ remains of order 1 in the limit $\alpha\to0$. In section \ref{sub:Empirical-validation-of}, we will see that this assumption is satisfied in the quasi-linear barotropic model.\\

In the regime $\alpha\ll1$, %if the condition $\tau_{\ell,\ell'}\sim1$ is fulfilled then 
we can consider a time $\Delta t$ much larger than the
auto-correlation times $\tau_{\ell,\ell'}$ but much smaller than
the typical time for the evolution of $x$ itself: $\tau_{\ell,\ell'}\ll \Delta t\ll1/\alpha$.
Over such time scale, (\ref{eq:slow-general}) can be integrated to
give 
\begin{equation}
x(t+\Delta t)=x(t)+\alpha\int_{t}^{t+\Delta t}f\left(x(u),y(u)\right)\mbox{d}u\simeq x(t)+\alpha\int_{t}^{t+\Delta t}f\left(x(t),\tilde{y}(u)\right)\mbox{d}u,\label{eq:slow-process-integrated}
\end{equation}
where in obtaining the last equality we have used the fact that over time
$\Delta t$ the process $x$ has almost not evolved. The relation (\ref{eq:slow-process-integrated})
is used in the following to derive equations for the average
behaviour, typical fluctuations and large fluctuations of $x$, in
the time scale separation limit $\alpha\ll1$.

\subsection{Average evolution and energy balance for the slow process\label{sub:Average-evolution-and}}

We now describe the typical dynamics of $x$ over time scales $\Delta t$
such that $\tau_{\ell,\ell'}\ll \Delta t\ll1/\alpha$, recovering classical
results from stochastic averaging \cite{Gardiner_1994_Book_Stochastic}.
Because the time $\Delta t$ in \eqref{eq:slow-process-integrated} is much
larger than the typical correlation time of the components of $f\left(x,\tilde{y}(u)\right)$,
by the Law of Large Numbers we can replace the time average by a statistical average: $\frac{1}{\Delta t}\int_{t}^{t+\Delta t}f\left(x,\tilde{y}(u)\right)\mbox{d}u\simeq F(x)$
where $F(x)\equiv\mathbb{E}_{x}\left[f\left(x,\tilde{y}(u)\right)\right]$
is the average force acting on $x$, computed in the statistically
stationary state of the virtual fast process (\ref{eq:virtual-fast-process}).
Then, the average evolution of $x$ at leading order in $\alpha \Delta t\ll1$
is 
\begin{equation}
\frac{\Delta x}{\Delta t}\equiv\frac{x(t+\Delta t)-x(t)}{\Delta t}\simeq\alpha F(x(t)).\label{eq:average-omegaz}
\end{equation}
In the case of zonal jet dynamics in barotropic models, $x$ is the
zonally averaged vorticity (or velocity) and $F(x)$ is the average advection term
$R$. The effective dynamics \eqref{eq:average-omegaz} is  very
close to S3T-CE2 types of closures \cite{marston65conover,bakas2015s3t,Srinivasan-Young-2011-JAS,tobias2013direct,marston2014direct}
or to kinetic theory \cite{Bouchet_Nardini_Tangarife_2013_Kinetic}.
This point is further discussed in section \ref{sub:Gaussian-approximation-of}.\\

The effective dynamics \eqref{eq:average-omegaz}
is not enough to describe the effective energy balance related to
the slow process $x$. Indeed, replacing the time averaged force in
\eqref{eq:slow-process-integrated} by its statistical average amounts
to neglecting fluctuations in the process $f(x,\tilde{y}(u))$. The
fluctuations are however relevant in the evolution of quadratic forms
of $x$. In particular, if we define the energy of the slow degrees
of freedom as $E=\frac{1}{2}x\cdot x=\sum_{\ell}E_{\ell}$ with $E_{\ell}=\frac{1}{2}x_{\ell}^{2}$,
an equation for $E_{\ell}$ can be derived using \eqref{eq:slow-process-integrated},
\begin{equation}\begin{aligned}
E_{\ell}(t+\Delta t)\simeq &\,E_{\ell}(t)+\alpha x_{\ell}(t)\int_{t}^{t+\Delta t}f_{\ell}\left(x(t),\tilde{y}(u)\right)\mbox{d}u\\
&+\frac{\alpha^{2}}{2}\int_{t}^{t+\Delta t}\int_{t}^{t+\Delta t}f_{\ell}\left(x(t),\tilde{y}\left(u_{1}\right)\right)f_{\ell}\left(x(t),\tilde{y}\left(u_{2}\right)\right)\mbox{d}u_{1}\mbox{d}u_{2}.\\
\end{aligned}\end{equation}
Define 
\begin{equation}
Z_{\ell,\ell'}(x)  \equiv\lim_{\Delta t\to\infty}\frac{1}{\Delta t}\int_{0}^{\Delta t}\int_{0}^{\Delta t}\mathbb{E}_{x}\left[\left[f_{\ell}\left(x,\tilde{y}\left(u_{1}\right)\right)f_{\ell'}\left(x,\tilde{y}\left(u_{2}\right)\right)\right]\right]\mbox{d}u_{1}\mbox{d}u_{2}\,,\label{eq:def-Xi-ell-ell}
\end{equation}
then using again that $\Delta t$ is much larger than the correlation time
of $f\left(x,\tilde{y}(u)\right)$ we get 
\begin{equation}
\frac{\Delta E_{\ell}}{\Delta t}\simeq\alpha x_{\ell}F_{\ell}(x)+\frac{\alpha^{2}}{2}Z_{\ell,\ell}(x).\label{eq:slow-energy-balance}
\end{equation}
This relation is the energy balance for the slow evolution of $x$:
$p_{mean,\ell}=\alpha x_{\ell}F_{\ell}(x)$ is the average energy
injection rate by the mean force $F(x)$, and $p_{fluct,\ell}=\frac{\alpha^{2}}{2}Z_{\ell,\ell}(x)$
is the average energy injection rate by the typical fluctuations of
the force $f$, as quantified by $Z(x)$. Neglecting the term $p_{fluct,\ell}$
in \eqref{eq:slow-energy-balance}, we recover the energy balance
we would have obtained by computing the evolution of $E_{\ell}$ from
\eqref{eq:average-omegaz}. This observation confirms the fact that
fluctuations of $f$, which are not taken into account in \eqref{eq:average-omegaz},
are relevant in the effective dynamics of $x$.

\subsection{Large Deviation Principle for the slow process\label{sub:Large-time-large-deviations}}

\subsubsection{Large deviation rate function for the action of the fast variable on the slow variable}

Equations \eqref{eq:average-omegaz} and \eqref{eq:slow-energy-balance}
give the evolution of $x$ and $x\cdot x$ at leading order in $\alpha\ll1$.
Such effective evolution equations can also be found in a more formal
way using stochastic averaging \cite{freidlin2012random,Gardiner_1994_Book_Stochastic}.
The effective equations only describe the low-order statistics of
the slow process:  The average evolution and typical fluctuations (variance
or energy). In contrast, the Large Deviation Principle gives access
to the statistics of both typical and rare events, also in the limit
$\alpha\ll1$.  For system (\ref{eq:slow-general}),
the Large Deviation Principle was first proved by Freidlin (see Ref. \onlinecite{freidlin2012random}
and references therein). It states that the probability density of
a path of the slow process $x$, denoted $\mathcal{P}[x]$, satisfies \cite{freidlin2012random}
\begin{equation}
\ln\mathcal{P}\left[x\right]\underset{\alpha\to0}{\sim}-\frac{1}{\alpha}\int\mathcal{L}\left(x(t),\dot{x}(t)\right)\mbox{d}t\label{eq:probability-path}
\end{equation}
with $\mathcal{L}\left(x,\dot{x}\right)\equiv\min_{\theta}\left\{ \dot{x}\cdot\theta-H\left(x,\theta\right)\right\} $
and where $H\left(x,\theta\right)$ is the scaled cumulant generating
function 
\begin{equation}
H\left(x,\theta\right)\equiv\lim_{\Delta t\to\infty}\frac{1}{\Delta t}\ln\mathbb{E}_{x}\left[\exp\left(\theta\cdot\int_{0}^{\Delta t}f\left(x,\tilde{y}(u)\right)\mbox{d}u\right)\right],\label{eq:SCGF-general}
\end{equation}
where we recall that $\mathbb{E}_{x}$ is an average over realisations
of the virtual fast process (\ref{eq:virtual-fast-process}) in its
statistically stationary state.  Quantities $H$ and $\mathcal{L}$
are classical definitions from Large Deviation Theory \cite{freidlin2012random}.
The knowledge of the function $H(x,\theta)$ is equivalent to the
knowledge of $\mathcal{L}\left(x,\dot{x}\right)$, which gives the
probability of any path of the slow process $x$ through (\ref{eq:probability-path}).
Computing $H\left(x,\theta\right)$ is thus a very efficient way to
study the effective statistics of $x(t)$, even when extremely rare events
that are not described in the effective equations \eqref{eq:average-omegaz}
and \eqref{eq:slow-energy-balance} play an important role.

Because the Large Deviation Principle \eqref{eq:probability-path} describes both rare events and typical events, information about the effective dynamics (\ref{eq:average-omegaz}, \ref{eq:slow-energy-balance}) is encoded in the definition of the scaled cumulant generating function. Indeed, a Taylor expansion in powers of $\theta$ in \eqref{eq:SCGF-general} gives
\begin{equation}
H\left(x,\theta\right)=\sum_{\ell}\theta_{\ell}F_{\ell}(x)+\frac{1}{2}\sum_{\ell,\ell'}\theta_{\ell}\theta_{\ell'}Z_{\ell,\ell'}(x)+O\left(\theta^{3}\right),\label{eq:SCGF-expansion-theta}
\end{equation}
with $F(x)\equiv\mathbb{E}_{x}\left[f\left(x,\tilde{y}(u)\right)\right]$
and $Z$ given by \eqref{eq:def-Xi-ell-ell}. The terms appearing
in the leading order evolution of $x$ \eqref{eq:average-omegaz}
and of the energy \eqref{eq:slow-energy-balance} are thus contained
in the scaled cumulant generating function, through \eqref{eq:SCGF-expansion-theta}.

Higher-order terms in (\ref{eq:SCGF-expansion-theta}) involve cubic
and higher-order cumulants of large time averages of the process $f\left(x,\tilde{y}(u)\right)$.
If this process is a Gaussian process,  its statistics are only
given by its first and second order cumulants \cite{Gardiner_1994_Book_Stochastic}.
As a consequence, for such process $H\left(x,\theta\right)$ is quadratic
in $\theta$ and (\ref{eq:SCGF-expansion-theta}) is exact (corrections
of order $\theta^{3}$ are exactly zero).\\

In practice, the scaled cumulant generating function (\ref{eq:SCGF-general})
involves the virtual fast process (\ref{eq:virtual-fast-process}).
This stochastic process depends only parametrically on $x$, which
means that we do not have to study the coupled system (\ref{eq:slow-general}) in order to compute $H(x,\theta)$.
This result is consistent with the time scale separation property
of (\ref{eq:slow-general}). In quasi-linear systems such as the quasi-linear
barotropic dynamics, the virtual fast process is an Ornstein-Uhlenbeck
process, which is particularly simple to study. This specific class
of systems is considered next in section \ref{sub:Quasi-linear-systems-with}.

\subsubsection{Quasi-linear systems with action of the fast process on the slow one through a quadratic force: the matrix Riccati equation\label{sub:Quasi-linear-systems-with}}

We are particularly interested in the more specific class of systems
defined by
\begin{equation}
\left\lbrace \begin{aligned} & \frac{dx}{dt}=\alpha y^{T}\mathcal{M}y+\alpha g\left(x\right)\\
 & \frac{dy}{dt}=-L_{x}\left[y\right]+\eta
\end{aligned}
\right.\label{eq:slow-quasilinear}
\end{equation}
 where $\mathcal{M}$ is a symmetric matrix, and $L_{x}$ is a linear
operator acting on $y$ that depends parametrically on $x$. The system
(\ref{eq:slow-quasilinear}) is a particular case of (\ref{eq:slow-general})
with $f(x,y)=y^{T}\mathcal{M}y+g\left(x\right)$ and $b\left(x,y\right)=-L_{x}\left[y\right]$.

When $x$ is the zonal flow vorticity profile and $y$ is the eddy
vorticity, the quasi-linear barotropic dynamics \eqref{eq:barotropic-quasi-linear}
is an example of such a system, where the quadratic form $y^{T}\mathcal{M}y$
defines the zonally averaged advection term $R$ and $g\left(x\right)$ contains the
dissipative terms acting on the large-scale zonal flow $x$, and where
$L_{x}$ is the linearized barotropic operator close to the zonal
flow $x$ (see also section \ref{sub:Application-of-the}).\\

We now describe the effective dynamics and large deviations of $x$
in the system (\ref{eq:slow-quasilinear}), in the limit $\alpha\to0$.
In this limit, the statistics of $y$ are very close to the statistics
of the virtual fast process \eqref{eq:virtual-fast-process}, which
in this case reads
\begin{equation}
\frac{d\tilde{y}}{dt}=-L_{x}\left[\tilde{y}\right]+\eta,\label{eq:Ornstein-Uhlenbeck}
\end{equation}
where $x$ is frozen. Equation \eqref{eq:Ornstein-Uhlenbeck} describes
an Ornstein-Uhlenbeck process, whose stationary distribution is Gaussian
\cite{Gardiner_1994_Book_Stochastic}. Then, the stationary statistics
of \eqref{eq:Ornstein-Uhlenbeck} are fully determined by the mean
and covariance of $\tilde{y}$. The mean is zero, and the covariance
$G_{ij}=\mathbb{E}\left[\tilde{y}_{i}\tilde{y}_{j}\right]$
is given by the Lyapunov equation
\begin{equation}
\frac{d G}{d t}+L_{x}G+GL_{x}^{T}=C.\label{eq:Lyapunov}
\end{equation}
The Lyapunov equation (\ref{eq:Lyapunov}) converges to a unique stationary solution whenever \eqref{eq:Ornstein-Uhlenbeck} has an invariant measure. We recall that such an invariant measure is required for the time scale separation hypothesis to be relevant. 
The effective dynamics of $x$ over times $\Delta t\ll1/\alpha$ is given
by \eqref{eq:average-omegaz}. In the case of (\ref{eq:slow-quasilinear}),
it reads
\begin{equation}
\frac{\Delta x}{\Delta t}\simeq\alpha\left[\mathcal{M}\cdot G_{\infty}(x)+g(x)\right]\label{eq:CE2}
\end{equation}
with $\mathcal{M}\cdot G_{\infty}(x)=\sum_{i,j}\mathcal{M}_{ij}\left(G_{\infty}\right)_{ij}(x)$
where $G_{\infty}$ is the stationary solution of the Lyapunov equation
\eqref{eq:Lyapunov}. 
Simulating the effective slow dynamics \eqref{eq:CE2}
can be done by integrating the Lyapunov equation \eqref{eq:Lyapunov},
using standard methods\footnote{The application ``GCM'' integrates the
equation \ref{eq:Lyapunov} and the effective dynamics \ref{eq:CE2}.}. 
It provides an effective description of the attractors of $x$,
and of the relaxation dynamics towards the attractors. Examples
of such numerical simulations of \eqref{eq:CE2} in the case of zonal
jet dynamics in the barotropic model can be found for instance in
Refs. \onlinecite{bakas2015s3t,Srinivasan-Young-2011-JAS,tobias2011astrophysical,tobias2013direct,marston2014direct}.\\

In order to describe large fluctuations of $x$ in (\ref{eq:slow-quasilinear}),
we need to use the Large Deviation Principle \eqref{eq:probability-path}.
In practice, we compute the scaled cumulant generating function
\eqref{eq:SCGF-general}. As proven in Ref. \onlinecite{BouchetTangarifeVandenEijnden2015}, for the system (\ref{eq:slow-quasilinear}),
the scaled cumulant generating function is given by
\begin{equation}
H\left(x,\theta\right)=\theta\cdot g(x)+\mbox{tr}\left(CN_{\infty}\left(x,\theta\right)\right)\label{eq:SCGF-quasilinear}
\end{equation}
where $C$ is the covariance matrix of the noise $\eta$ in (\ref{eq:slow-quasilinear})
and $N_{\infty}\left(x,\theta\right)$ is a symmetric matrix, stationary
solution of 
\begin{equation}
\frac{d N}{d t}+NL_{x}+L_{x}^{T}N=2NCN+\theta\mathcal{M}.\label{eq:NL-Lya}
\end{equation}
Equation (\ref{eq:NL-Lya})
is a particular case of a matrix Ricatti equation, and in the following
we refer to (\ref{eq:NL-Lya}) as the Ricatti equation. 
$\theta$ is the parameter of the cumulant generating function \eqref{eq:SCGF-general} that defines $H$. Whenever $\theta$ is in the parameter range for which the limit in \eqref{eq:SCGF-general} exists, called the admissible $\theta$ range, Eq. (\ref{eq:NL-Lya}) has a stationary solution. For the case in this section, with a linear dynamics with a quadratic observable, the admissible $\theta$ range is easily studied through the analysis of the positivity of a quadratic form. One can conclude that the admissible $\theta$ range is an interval containing $0$. All the information regarding the large deviation rate function is contained in the values of $H$ for $\theta$ in this range. 

The Ricatti equation (\ref{eq:NL-Lya}) is similar to the Lyapunov equation
\eqref{eq:Lyapunov}, and it can be solved using similar methods\footnote{Note that the ordering of products with $L_{x}$ and $L_{x}^{T}$
differs between \eqref{eq:Lyapunov} and \eqref{eq:NL-Lya}.}. Moreover, the numerical implementation of (\ref{eq:SCGF-quasilinear},
\ref{eq:NL-Lya}) can be easily checked using the relation with the
Lyapunov equation \eqref{eq:Lyapunov}. Namely, \eqref{eq:SCGF-expansion-theta}
implies that
\[
\left.\frac{dH}{d\theta}\right|_{\theta=0}=\mathcal{M}\cdot G_{\infty}(x)+g(x).
\]
The first term in the right-hand side is computed from the Lyapunov
equation \eqref{eq:Lyapunov}, while the left-hand side is computed
from the Ricatti equation (\ref{eq:NL-Lya}) together with~(\ref{eq:SCGF-quasilinear}).

In section \ref{sub:Application-of-the}, we present a numerical
resolution of (\ref{eq:NL-Lya}) for the case of the quasi-linear
barotropic equation on the sphere, and  compute directly the scaled
cumulant generating function using (\ref{eq:SCGF-quasilinear}). We
show that (\ref{eq:NL-Lya}) can be very easily solved for a given
value of $\theta$. This means that the result (\ref{eq:SCGF-quasilinear})
permits the study of arbitrarily rare events in zonal jet dynamics extremely
easily, through the Large Deviation Principle \eqref{eq:probability-path}.
Such result is in clear contrast with approaches through direct numerical
simulations, which require that the total time of integration increases
as the probability of the event of interest decreases. This limitation
of direct numerical simulations in the study of rare events statistics
is made more precise in next section.

%%%%%%%%%%%%%%%%%%%%%%%%%%%%%%%%
%%%%%%%%%%%%%%%%%%%%%%%%%%%%%%%%
%%%%%%%%%%%%%%%%%%%%%%%%%%%%%%%%
\subsection{Estimation of the large deviation function from time series analysis\label{sec:LD-estimation-SCGF}}

In this section we present a way to compute the scaled cumulant generating
function (\ref{eq:SCGF-general}) from a time series of the virtual
fast process \eqref{eq:virtual-fast-process}, for instance one obtained
from a direct numerical simulation. Many of the technical aspects of this empirical approach follow Ref. \onlinecite{rohwer2014convergence}.

Consider a time series $\left\{ \tilde{y}(u)\right\} _{0\leq u\leq T}$
of the virtual fast process \eqref{eq:virtual-fast-process}, with
a given total time window $u\in[0,T]$. Because the quantities of interest like $H(x,\theta)$ involve expectations in the stationary state of the virtual fast process, we assume that the time series $\left\{ \tilde{y}(u)\right\} _{0\leq u\leq T}$ corresponds to this stationary state. We use the continuous time series notation for simplicity. The generalization of the following formulas to the case of discrete time series is straightforward. For simplicity, we also denote by $R(u)\equiv f\left(\tilde{y}(u)\right)$, the quantity for which the scale cumulant generating function $H\left(\theta\right)=\lim_{t\rightarrow \infty}\frac{1}{t}\log\mathbb{E}\exp\left(\theta\int_{0}^{t}R(u)\,\mbox{d}u\right)$ should be estimated.

The basic method to estimate the scaled cumulant generating function
(\ref{eq:SCGF-general}) is to divide the full time series $\left\{ \tilde{y}(u)\right\} _{0\leq u\leq T}$
into blocks of length $\Delta t$, to compute the integrals $\int_{t_{0}}^{t_{0}+\Delta t}R(u)\,\mbox{d}u$
over those blocks, and  to average the quantity $\exp\left(\theta\cdot\int_{t_{0}}^{t_{0}+\Delta t}R(u)\,\mbox{d}u\right)$.
For a small block length $\Delta t$, the large-time regime defined by the
limit $\Delta t\to\infty$ in the theoretical expression of $H$ (\ref{eq:SCGF-general})
is not attained. On the other hand, too large values of $\Delta t$ ---typically
of the order of the total time $T$--- lead to a low number of blocks,
and thus to a very poor estimation of the empirical mean. Estimating
$H$ thus requires finding an intermediate regime for $\Delta t$. More precisely,
we expect this regime to be attained for $\Delta t$ equal to a few times
the correlation time of the process $R(u)$, defined by \cite{newman1999monte,papanicolaou1977introduction}
\begin{equation}
\tau \equiv \lim_{\Delta t\to\infty} \frac{\int_0^{\Delta t}\int_0^{\Delta t} \mathbb{E}_z\left[\left[\,R(u_1) R(u_2)\,\right] \right] \,\mathrm{d} u_1\mathrm{d} u_2}{2\Delta t\,\mathbb{E}_z\left[\left[\,R^2\,\right] \right]}  = \frac{\int_0^\infty \mathbb{E}_z\left[\left[\,R(u) R(0)\,\right] \right] \,\mathrm{d} u}{\mathbb{E}_z\left[\left[\,R^2\,\right] \right]}\,,
\label{eq:LD-tau_corr-def}
\end{equation}
where $\mathbb{E}_z[[R(u_1) R(u_2)]]$ is the covariance of $R$ at time $u_1$ and at time $u_2$. The second equality is easily obtained assuming that the process $R(u)$ is stationary. Because of the infinite-time
limit in \eqref{eq:LD-tau_corr-def}, the estimation of $\tau$ suffers
from the same finite sampling problem as the estimation of $H$.

Finding a block length $\Delta t$
such that the estimation of $H$ and $\tau$ is accurate is thus a
tricky problem. In the following, we propose a method to find the
optimal $\Delta t$ and estimate the quantities we are interested in. The
proposed method is close to the ``data bunching'' method used to
estimate errors in Monte Carlo simulations \cite{krauth2006statistical}.

%%%%%%%%%%%%%%%%%%%%%%%%%%%%%%%%
\subsubsection{Estimation of the correlation time\label{sub:Estimation-of-the}}

We first consider the problem of the estimation of $\tau$ in
a simple solvable case, so the numerical results can be compared directly
to explicit formulas. Consider the stochastic process $R=w^{2}$
where $w$ is the one-dimensional Ornstein-Uhlenbeck process 
\begin{equation}
\frac{dw}{dt}=-w+\eta,\label{eq:OU-1D}
\end{equation}
where $\eta$ is a Gaussian white noise with correlation $\mathbb{E}\left(\eta(t)\eta(t')\right)=\delta(t-t')$. A direct calculation gives the correlation time of $R$, $\tau=1/2$. Using \eqref{eq:SCGF-quasilinear} and \eqref{eq:NL-Lya}, the scaled cumulant generating function can also be computed explicitly (see for instance Ref. \onlinecite{BouchetTangarifeVandenEijnden2015}). We obtain
\begin{equation}
H(\theta) = \frac12 - \frac12 \sqrt{1-2\theta},
\label{eq:exact-SCGF}
\end{equation}
defined for $\theta\leq 1/2$.

For a time series $\left\{ R(u)\right\} _{0\leq u\leq T}$, we denote by
$\bar{R}_{T}=\frac{1}{T}\int_{0}^{T}R(u)\,\mbox{d}u$ and by $\mbox{var}_{T}(R)=\frac{1}{T}\int_{0}^{T}\left(R(u)-\bar{R}_{T}\right)^{2}\mbox{d}u$
respectively the empirical mean and variance of $R$ over the full
time series. We  estimate the correlation time $\tau$ defined
in (\ref{eq:LD-tau_corr-def}) using an average over blocks
of length $\Delta t$, 
\begin{equation}
\tau_{\Delta t}=\frac{1}{2\Delta t\,\mbox{var}_{T}(R)}\mathbb{E}_{\frac{T}{\Delta t}}\left[\left(\int_{t_{0}}^{t_{0}+\Delta t}\left(R(u)-\bar{R}_{T}\right)\,\mbox{d}u\right)^{2}\right],\label{eq:tau-estimation}
\end{equation}
where $\mathbb{E}_{\frac{T}{\Delta t}}\left[h_{t_{0}}\right]$ is the empirical
average over realisations of the quantity $h_{t_{0}}$ inside the
brackets\footnote{Explicitely, 
\begin{equation}
\mathbb{E}_{\frac{T}{\Delta t}}\left[\left(\int_{t_{0}}^{t_{0}+\Delta t}\left(R(s)-\bar{R}_{T}\right)\,\mbox{d}s\right)^{2}\right]=\frac{\Delta t}{2T}\sum_{k=0}^{\frac{2T}{\Delta t}-2}\left(\int_{k\Delta t/2}^{k\Delta t/2+\Delta t}\left(R(u)-\bar{R}_{T}\right)\,\mbox{d}u\right)^{2}\,,\label{eq:empirical-average-welch}
\end{equation}
assuming for simplicity that $T/\Delta t$ is an integer. Generalisations
to any $T,\Delta t$ is straightforward, replacing $2T/\Delta t$ by its floor value.
The 50\% overlap is suggested by Welch's estimator of the power spectrum
of a random process \cite{welch1967use}.}.

\begin{figure}\begin{center}
\includegraphics[scale=0.5]{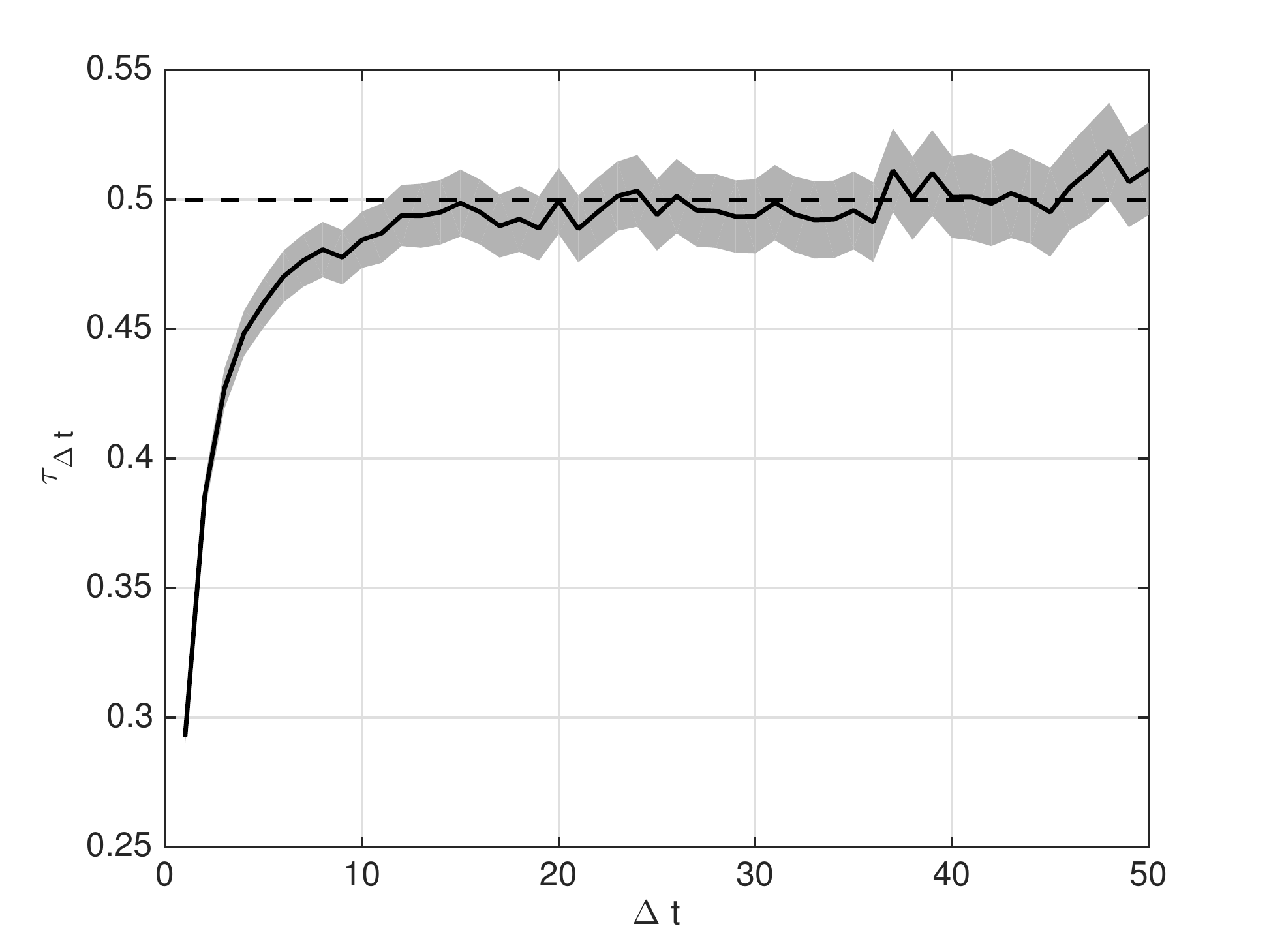}
\caption{\label{fig:estimation-tau}Plot of the estimated correlation time
$\tau_{\Delta t}$ (black line) and error bars (grey shading) as functions
of $\Delta t$.
For small values of $\Delta t$, the large-time limit in (\ref{eq:LD-tau_corr-def})
is not achieved, which explains the low values of $\tau_{\Delta t}$. For
too large values of $\Delta t$, the empirical average $\mathbb{E}_{\frac{T}{\Delta t}}$
in (\ref{eq:tau-estimation}) is not accurate due to the small value
of $\frac{T}{\Delta t}$, which explains the increasing fluctuations in $\tau_{\Delta t}$
as $\Delta t$ increases. The optimal value $\Delta t^{\star}$ is the one in between
these artificial behaviour. Here, one can read $\Delta t^{\star}\simeq20$
and $\tau_{\Delta t^{\star}}\simeq0.5$, in agreement with the exact value
$\tau=1/2$ (dashed line). The Ornstein-Uhlenbeck
process (\ref{eq:OU-1D}) has been integrated over $T=5.10^{4}$ using the method proposed in Ref. \onlinecite{Gillespie1996}, with time step $10^{-3}$.}
\end{center}\end{figure}

To find the optimal value of $\Delta t$, we plot $\tau_{\Delta t}$ as a function
of $\Delta t$ in figure \ref{fig:estimation-tau}. For small values of
$\Delta t$, the large-time limit in (\ref{eq:LD-tau_corr-def})
is not achieved, which explains the low values of $\tau_{\Delta t}$. For
too large values of $\Delta t$, the empirical average $\mathbb{E}_{\frac{T}{\Delta t}}$
in \eqref{eq:tau-estimation} is not accurate due to the small value
of $\frac{T}{\Delta t}$ (small number of blocks), which explains the increasing fluctuations in $\tau_{\Delta t}$
as $\Delta t$ increases. The optimal value of $\Delta t$ ---denoted $\Delta t^{\star}$
in the following--- is between the values giving these artificial behaviours.
It should satisfy $T\gg \Delta t^{\star}\gg\tau_{\Delta t^{\star}}$. Here, one
can read $\Delta t^{\star}\simeq10$ and $\tau_{\Delta t^{\star}}\simeq0.5$, so
this optimal $\Delta t^{\star}$ satifies the aforementioned condition. The
estimated value $\tau_{\Delta t^{\star}}$ is in agreement with the theoretical
value $\tau=1/2$.

The error bars for $\tau_{\Delta t}$ are given by $\Delta\tau_{\Delta t}=\sqrt{\mbox{var}\left(\tau_{\Delta t}\right)/N_{terms}}$,
where $\mbox{var}\left(\tau_{\Delta t}\right)$ is the empirical variance
associated with the average $\mathbb{E}_{\frac{T}{\Delta t}}$ defined in
(\ref{eq:empirical-average-welch}), and $N_{terms}$ is the number
of terms in this sum (roughly $N_{terms}\simeq2T/\Delta t$).

%%%%%%%%%%%%%%%%%%%%%%%%%%%%%%%%
\subsubsection{Estimation of the scaled cumulant generating function\label{sub:Sampling-the-SCGF}}

The self-consistent estimation of the correlation time $\tau$
presented in the previous section gives the optimal value $\Delta t^{\star}$
of the block length. Then, the scaled cumulant generating function
is computed for a given value of $\theta$ as

\begin{equation}
H_{T}\left(\theta\right)\equiv\frac{1}{\Delta t^{\star}}\ln\mathbb{E}_{\frac{T}{\Delta t^{\star}}}\left[\exp\left(\theta\int_{t_{0}}^{t_{0}+\Delta t^\star}R(u)\,\mbox{d}u\right)\right],\label{eq:SCGF-empirical}
\end{equation}
where $\mathbb{E}_{\frac{T}{\Delta t}}$ is the empirical average over the
blocks, as defined in \eqref{eq:empirical-average-welch}. However, the knowledge of $H\left(x,\theta\right)$ for an arbitrarily large value
of $\left|\theta\right|$ leads to the probability of an arbitrarily
rare event for the slow process $x$ through the Large Deviation Principle \eqref{eq:probability-path}. This is in contradiction with the fact that the
available time series $\left\{ R(u)\right\} _{0\leq u\leq T}$ is
finite. In other words, the range of values of $\theta$ for which
the scaled cumulant generating function $H_{T}(\theta)$ can be computed
with accuracy depends on $T$.

Indeed, for large positive values of $\theta$, the sum $\mathbb{E}_{\frac{T}{\Delta t^{\star}}}$
in (\ref{eq:SCGF-empirical}) is dominated by the largest term $\exp\left(\theta  I_{max}\right)$
where $I_{max}=\max_{t_{0}}\left\{ \int_{t_{0}}^{t_{0}+\Delta t}R(u)\,\mbox{d}u\right\} $
is the largest value of $\int_{t_{0}}^{t_{0}+\Delta t}R(u)\,\mbox{d}u$ over
the finite sample $\left\{ R(u)\right\} _{0\leq u\leq T}$. Then $H_{T}(\theta)\sim\frac{1}{\Delta t^{\star}}I_{max}\theta$
for $\theta\gg1$. This phenomenon is known as linearization \cite{rohwer2014convergence},
and is clearly an artifact of the finite sample size. We denote by $\theta_{max}$
the value of $\theta$ such that linearization occurs for $\theta>\theta_{max}$
. Typically, we expect $\theta_{max}$ to be a positive increasing
function of $T$. The same way, $H_{T}(\theta)\sim-\frac{1}{\Delta t^{\star}}I_{min}\theta$
for $\theta<0$ and $\left|\theta\right|\gg1$, with $I_{min}=\min_{t_{0}}\left\{ \int_{t_{0}}^{t_{0}+\Delta t}R(u)\,\mbox{d}u\right\} $.
In a similar way, we define $\theta_{min}$ as the minimum value of $\theta$ for which linearization occurs. Typically, we expect $\theta_{min}$ to
be a negative decreasing function of $T$.

\begin{figure}\begin{center}
\includegraphics[height=7cm]{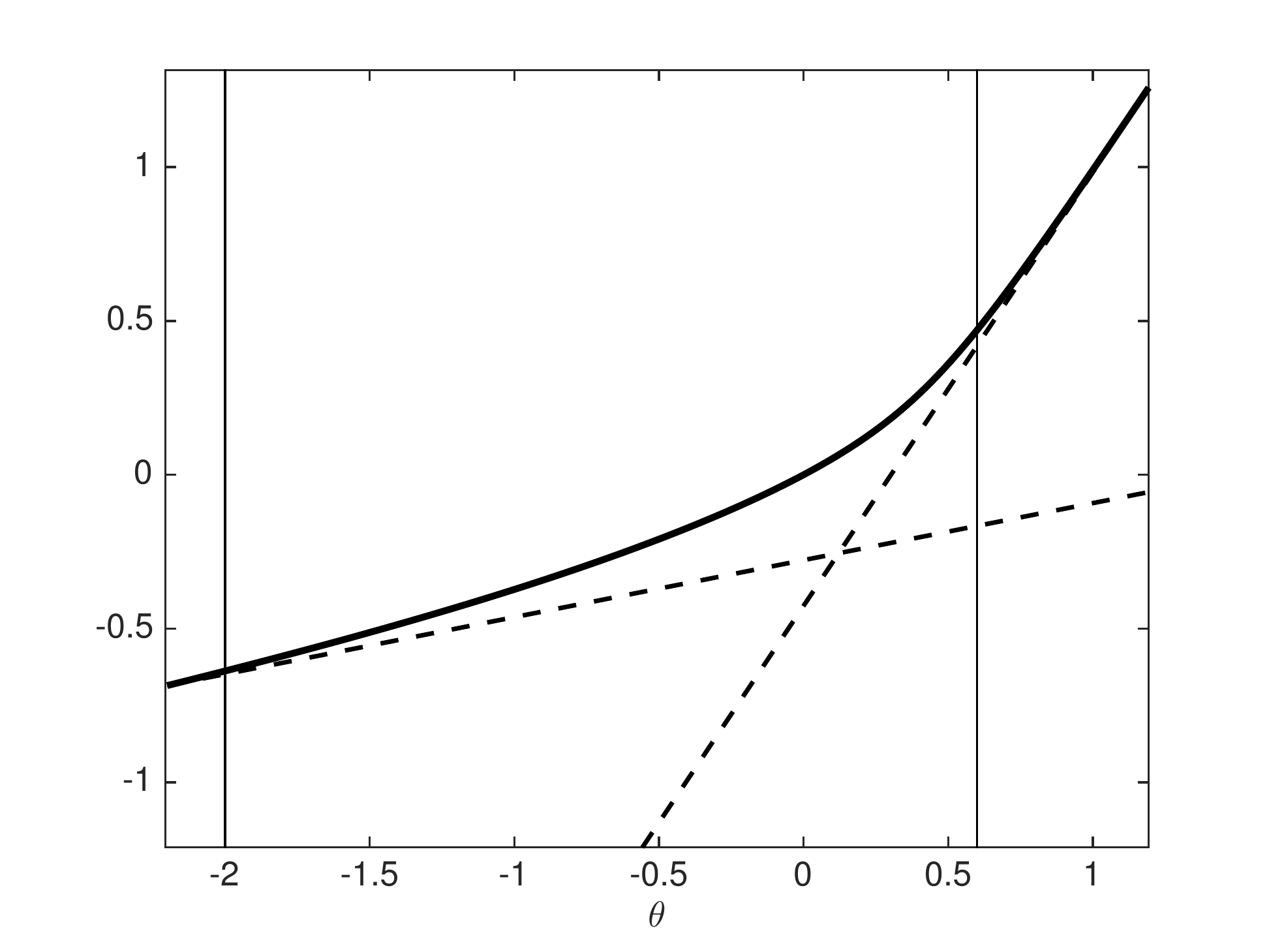}\\
\includegraphics[height=7cm]{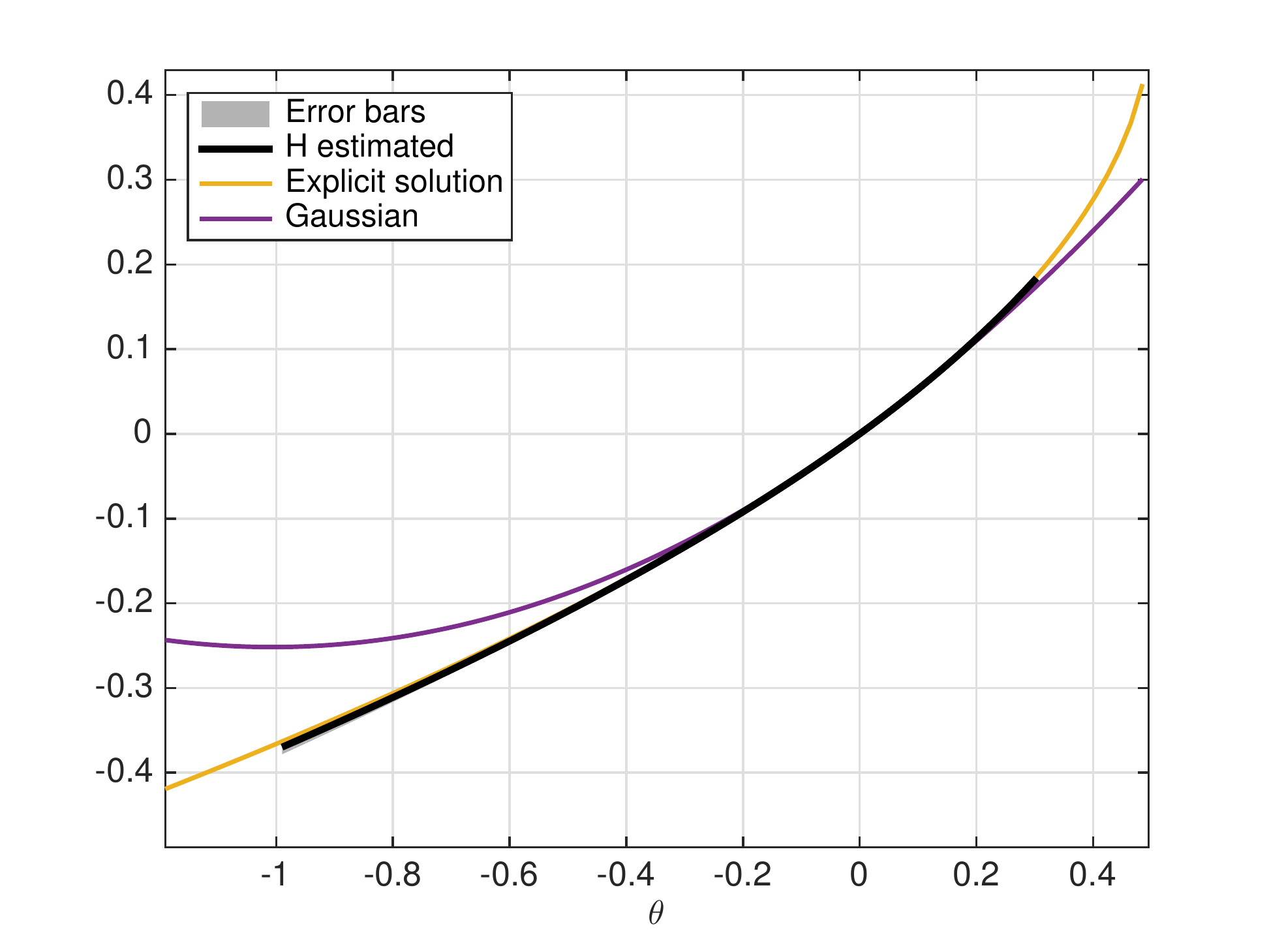}
\caption{\label{fig:SCGF-estimation}Computation of the scaled cumulant generating
function from (\ref{eq:SCGF-empirical}) for the one-dimensional Ornstein-Uhlenbeck
process (\ref{eq:OU-1D}). Upper panel: illustration of the linearization effect
for large values of $\left|\theta\right|$. The solid curve is the estimated
scaled cumulant generating function $H_T$, and the dashed lines
are the expected linear tails, which are artifacts of the finite sample
size \cite{rohwer2014convergence}. The thin vertical lines show the
range $\theta\in\left[\theta_{min},\theta_{max}\right]$ for which
we consider that linearization does not take place. Bottom pannel: the
converged scaled cumulant generating function estimator $H_{T}$ on
$\theta\in\left[\theta_{min}/2,\theta_{max}/2\right]$ (thick black
curve, with error bars in grey shading). The yellow curve is the exact
scaled cumulant generating function \eqref{eq:exact-SCGF}, it
fits the estimated one within statistical errors. The purple curve
is the quadratic approximation, that corresponds to a Gaussian process $R(u)$
(see equation \eqref{eq:SCGF-expansion-theta}). This quadratic approximation
is computed using the exact
mean, variance and correlation time of $R$. The Ornstein-Uhlenbeck
process (\ref{eq:OU-1D}) has been integrated over $T=5 \times10^{4}$ using the method proposed in Ref. \onlinecite{Gillespie1996}, with time step $10^{-3}$.}
\end{center}\end{figure}

The convergence of estimators like (\ref{eq:SCGF-empirical}) is studied
in Ref. \onlinecite{rohwer2014convergence}, in particular it is shown that error
bars can be computed in the range $\left[\theta_{min}/2,\theta_{max}/2\right]$
for a given time series $\left\{ R(u)\right\} _{0\leq u\leq T}$.
An example of a computation of $H_{T}(\theta)$ is shown in Figure \ref{fig:SCGF-estimation}
for the one-dimensional Ornstein-Uhlenbeck process, and compared to
the explicit solution. The full error bars
in Figure \ref{fig:SCGF-estimation} are given by the error from the
estimation of $\tau$ and the statistical error described in Ref. \onlinecite{rohwer2014convergence}.
The method shows excellent agreement with theory, and exposes non-Gaussian behavior.\\

In sections \ref{sub:Gaussian-approximation-of} and \ref{sub:Application-of-the}, we apply the tools (estimation of the correlation time and of the scaled cumulant generating function) to study the statistics of Reynolds' stresses in zonal jet dynamics.

%%%%%%%%%%%%%%%%%%%%%%%%%%%%%%%%
%%%%%%%%%%%%%%%%%%%%%%%%%%%%%%%%
%%%%%%%%%%%%%%%%%%%%%%%%%%%%%%%%
\section{Zonal energy balance and time scale separation in the inertial limit\label{sub:Gaussian-approximation-of}}

In this section we discuss the effective evolution and effective energy balance for zonal flows in the inertial regime $\nu_n\ll\alpha\ll1$, using the general results of section \ref{sub:Average-evolution-and} and numerical simulations.

\subsection{Effective dynamics and energy balance for the zonal flow\label{sub:energy-balance}}

Using \eqref{eq:zonal-vorticity-ell} and \eqref{eq:average-omegaz}, the effective evolution of the zonal jet velocity profile $U(\phi,t)$ in the regime $\nu_n\ll\alpha\ll1$ reads
\begin{equation}
\frac{\partial U}{\partial t} \simeq\alpha F[U] - \alpha U - \nu_n (-\Delta)^n U,
\label{eq:LD-effective-U}
\end{equation}
with $F[U]\equiv \mathbb{E}_U[f]$ where $f$ is minus the Reynolds' stress divergence and $\mathbb{E}_U$ is the average in the statistically stationary state of the linear barotropic dynamics \eqref{eq:frozen-QL}, with $U$ held fixed.

Equation \eqref{eq:LD-effective-U} describes the effective slow dynamics of zonal jets in the regime $\nu_n\ll\alpha\ll1$, it is the analogous of the kinetic equation proposed in Ref. \onlinecite{Bouchet_Nardini_Tangarife_2013_Kinetic}. In particular, the attractors of \eqref{eq:LD-effective-U} are the same as the attractors of a second order closure of the barotropic dynamics \cite{marston65conover,ait2015cumulant}.\\

As explained in a general setting in section \ref{sub:Average-evolution-and}, equation  \eqref{eq:LD-effective-U} only takes into account the average Reynolds' stresses (through the term $F[U]$).  As a consequence it does not describe accurately the effective zonal energy balance. Quantifying the influence of fluctuations of Reynolds' stresses on the zonal energy balance is one of the goals of this study. We now derive the effective zonal energy balance, and  describe the relative influence of average and fluctuations of Reynolds' stresses using numerical simulations.

\begin{figure}\begin{center}
\includegraphics[height=8cm]{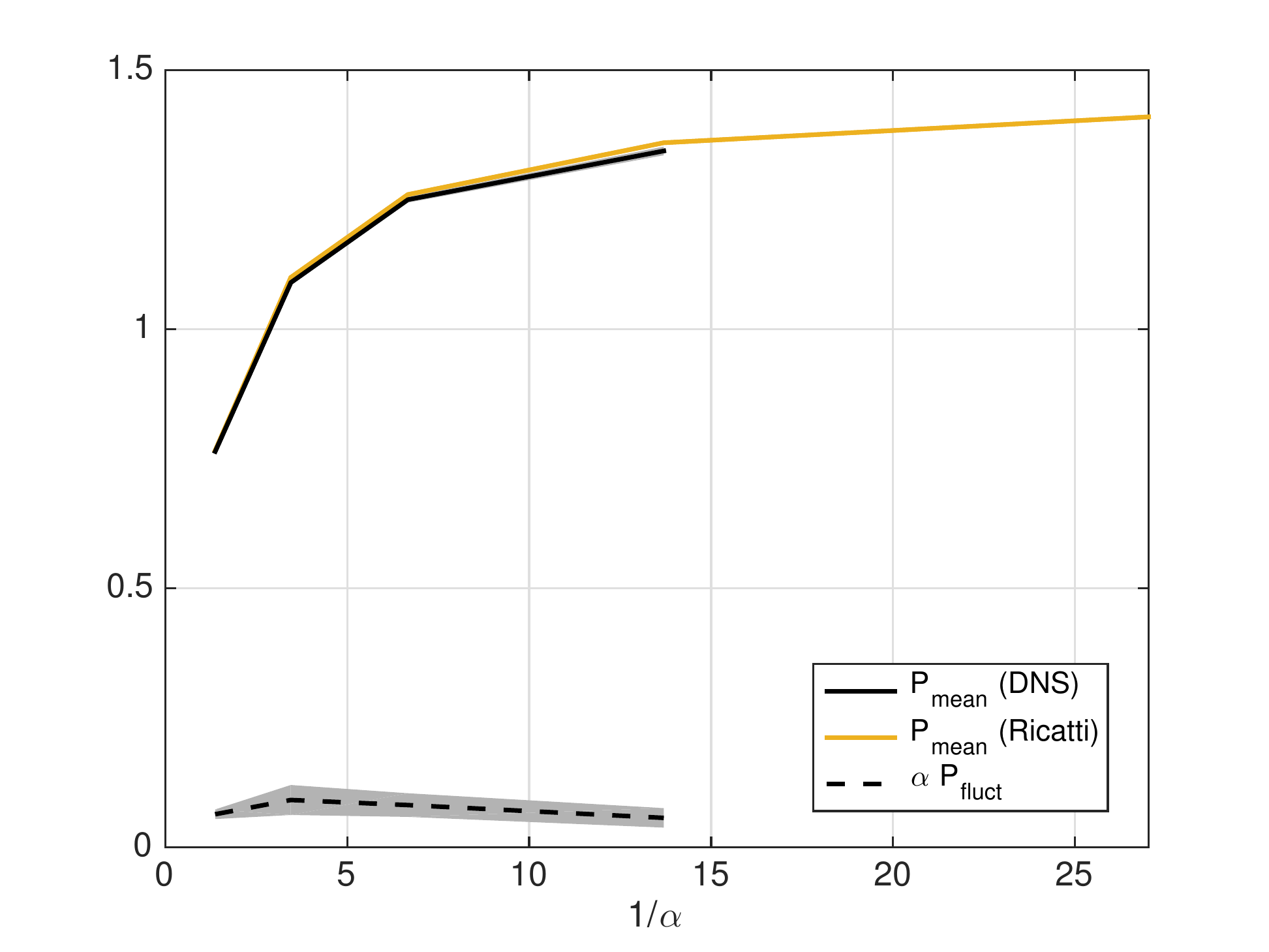}
\caption{\label{fig:energy-balance-total}Total energy injection rate into
the zonal flow by the mean Reynolds' stresses $P_{mean}$ (first term
in the r.h.s of (\ref{eq:zonal-energy-balance-total}), in solid line)
and by the fluctuations of Reynolds' stresses $\alpha P_{fluct}$
(last term in the r.h.s of (\ref{eq:zonal-energy-balance-total}),
in dashed line with statistical error bars in grey shading) as a function of $1/\alpha$. The quantities are
estimated from direct numerical simulations (DNS) of the linearized barotropic
equation \eqref{eq:virtual-fast-barotropic-QL} with parameters given in section \ref{sub:Numerical-implementation}, and $P_{mean}$ is also computed directly using the Ricatti equation \eqref{eq:NL-Lya} (yellow curve). %This allows to use finer resolution and smaller viscosity very easily, here the spectral cutoff in the Ricatti calculation is $L=120$ (compared to $L=80$ for the DNS), and the hyper-viscosity coefficient is such that the smallest scale has a damping rate of 4 (i.e. it is half of the hyperviscosity coefficient in the case $L=80$). The comparison of the solid black and yellow curves indicates that numerical resolution and hyper-viscosity are negligible in the computation of $P_{mean}$. We observe that $P_{mean}$ is of the same order as the zonal energy dissipation rate due to linear friction (second term in the r.h.s of (\ref{eq:zonal-energy-balance-total})), and that $\alpha P_{fluct}$ is about an order of magnitude smaller. Neglecting $\alpha P_{fluct}$  in \eqref{eq:zonal-energy-balance-total} leads to an error in the zonal energy budget of about 5--10\%. Besides, $P_{mean}$ increases up to a finite value as $\alpha\to0$, in agreement with theoretical predictions.
}
\end{center}\end{figure}

First note that the
hyperviscous terms in \eqref{eq:zonal-vorticity-ell} essentially
dissipate energy at the smallest scales of the flow. In the turbulent
regime we are interested in, such small-scale dissipation is negligible
in the global energy balance. For this reason, the
viscous terms can be neglected in \eqref{eq:LD-effective-U} and in the zonal energy balance. Note however that some hyper-viscosity is still present in the numerical simulations of the linear barotropic equation (\ref{eq:frozen-QL}), in order to ensure numerical stability. For consistency, we make sure that the hyper-viscous terms do not influence the numerical results, see Figure \ref{fig:energy-balance-total}.

The kinetic energy contained in zonal degrees of freedom reads $E_{z}=\int\mbox{d}\phi\,E\left(\phi\right)$
with $E\left(\phi\right)=\pi\cos\phi\,U^{2}\left(\phi\right)$. Using \eqref{eq:slow-energy-balance} we get the equation for the effective evolution of $E\left(\phi\right)$: 
\begin{equation}
\frac{1}{\alpha}\frac{dE}{dt}=p_{mean}(\phi)-2E+\alpha p_{fluct}(\phi)\,.\label{eq:energy-balance-kinetic}
\end{equation}
The left hand side is the instantaneous energy injection rates into the zonal mean
flow. It is equal to the sum of the average Reynolds' stresses $p_{mean}\left(\phi\right)\equiv2\pi\cos\phi\,F[U]\left(\phi\right)U\left(\phi\right)$, $-2E$, and the fluctuations of Reynolds' stresses $\alpha p_{fluct}\left(\phi\right)\equiv\alpha\pi\cos\phi\,Z[U]\left(\phi\right)$, where
\begin{equation}
Z[U]\left(\phi\right)\equiv \lim_{\Delta t\to\infty}\frac{1}{\Delta t}\int_{0}^{\Delta t}\int_{0}^{\Delta t}\mathbb{E}_{U}\left[\left[f\left(\phi,u_{1}\right)f\left(\phi,u_{2}\right)\right]\right]\mbox{d}u_{1}\mbox{d}u_{2}\,.
\end{equation}
Integrating \eqref{eq:energy-balance-kinetic} over latitudes, we
obtain the total zonal energy balance
\begin{equation}
\frac{1}{\alpha}\frac{dE_{z}}{dt}=P_{mean}-2E_{z}+\alpha P_{fluct},\label{eq:zonal-energy-balance-total}
\end{equation}
with $P_{mean}\equiv\int\mbox{d}\phi\,p_{mean}(\phi)$ and $\alpha P_{fluct}\equiv\int\mbox{d}\phi\,\alpha p_{fluct}(\phi)$.

All the terms appearing in \eqref{eq:energy-balance-kinetic} and \eqref{eq:zonal-energy-balance-total} can be easily estimated using data from a direct
numerical simulation of the linearized barotropic equation \eqref{eq:frozen-QL}.
Indeed, $F[U](\phi)$ can be computed as the empirical average of $f(\phi)$ in the stationary state
of \eqref{eq:frozen-QL}, and $Z[U](\phi)$ can be computed
using the method described in section \ref{sub:Estimation-of-the}
to estimate correlation times\footnote{The statistical error bars for $p_{fluct}$ are computed from the
error in the estimation of $Z[U](\phi)$, which is similar to the estimation
of the correlation time $\tau$ described in section \ref{sub:Estimation-of-the}.
The statistical error bars for $p_{mean}$ are computed from the error
in the estimation of the average $F$, given by $(\delta F)^{2}=\frac{1+2\tau/\Delta t}{N}\mbox{var}(F)$
where $\tau$ is the autocorrelation time of $F$, $\Delta t$ the
time step between measurements of the Reynolds' stress and $N$ the
total number of data points \cite{newman1999monte}.}.

The functions $F[U]$ and $Z[U](\phi)$ may be computed directly from the scaled cumulant generating function $H$, using \eqref{eq:SCGF-expansion-theta}. Computing $H$ using the Ricatti equation (\ref{eq:SCGF-quasilinear}, \ref{eq:NL-Lya}) and  using \eqref{eq:SCGF-expansion-theta}, we have a very easy way to compute the terms appearing in the effective slow dynamics \eqref{eq:LD-effective-U} or in the zonal energy balance equations \eqref{eq:energy-balance-kinetic} and \eqref{eq:zonal-energy-balance-total}, without having to simulate directly the fast process~\eqref{eq:frozen-QL}.\\

We now describe the results obtained by solving numerically the linearized barotropic equation \eqref{eq:frozen-QL}, where the mean flow velocity, $U$, is obtained from a quasilinear simulation as described in the end of section \ref{sub:Numerical-implementation}, and represented in Figure \ref{fig:U}. The energy injection rates $P_{mean}$ and $\alpha P_{fluct}$, computed using both of the methods explained above, with different values of the non-dimensional damping rate $\alpha$ are represented
in Figure \ref{fig:energy-balance-total}. The first term $P_{mean}$
(solid curve) is roughly of the order of magnitude of the dissipation
term in \eqref{eq:zonal-energy-balance-total} (recall we use units
such that $E_{z}\simeq1$). The second term $\alpha P_{fluct}$ is
about an order of magnitude smaller than $P_{mean}$. In this case, the energy balance \eqref{eq:zonal-energy-balance-total} implies that the zonal velocity is actually slowly decelerating.

Here, neglecting $\alpha P_{fluct}$ in \eqref{eq:zonal-energy-balance-total} leads to an error in the zonal energy budget of about 5--10\%. This confirms the fact that fluctuations of Reynolds' stresses are only negligible in a first approximation, and that they should be taken into account in order to obtain a quantitative description of zonal jet evolution. However, we emphasize that only one mode is stochastically forced in this case (see section \ref{sub:Numerical-implementation} for details). When several modes are forced independently, the Reynolds' stress divergence $f(\phi)$ is computed as the sum of independent contributions from each mode. If the number $K$ of forced modes becomes large, then the Central Limit Theorem implies that the typical fluctuations of $f(\phi)$ (and thus $\alpha P_{fluct}$) roughly scale as $1/K$. In Figure \ref{fig:energy-balance-total}, $K=1$ so we are basically considering the case where fluctuations of Reynolds' stresses are the most important in the zonal energy balance. In other words, this is the worst case test for CE2 types of closures. In most previous studies of second order closures like CE2, a large number of modes is forced \cite{marston2010statistics,tobias2013direct}, so in these cases $p_{fluct}(\phi)$ and $\alpha P_{fluct}$ are most likely to be negligible in the zonal energy balance.

We also observe that $P_{mean}$ increases up to a finite value as $\alpha\ll1$, while $\alpha P_{fluct}$ is nearly constant over the range of values of $\alpha$ considered. We further comment the behavior in the following.\\

\begin{figure}\begin{center}
\includegraphics[scale=0.5]{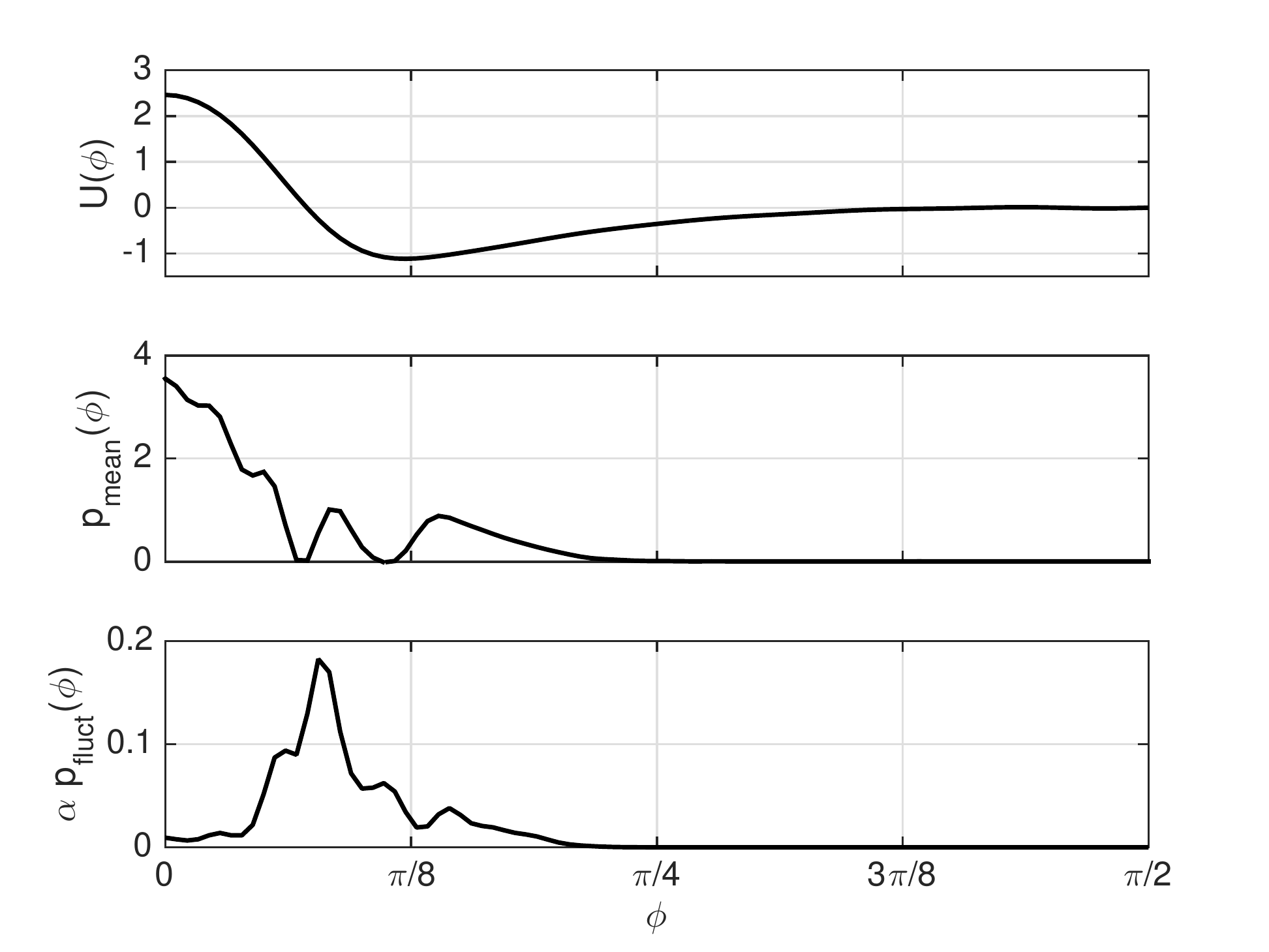}

\caption{\label{fig:comparison-U-p}From top to bottom: zonal velocity profile
$U(\phi)$, energy injection rate by the average Reynolds' stresses
$p_{mean}(\phi)$ and energy injection rate by the fluctuations of
Reynolds' stresses $\alpha p_{fluct}(\phi)$, as functions of latitude
$\phi$ restricted to the northern hemisphere. The values in the southern
hemisphere are symmetric with respect to northern hemisphere, see
Figures \ref{fig:U}, \ref{fig:energy-balance-mean} and \ref{fig:energy-balance-fluct}. $p_{mean}$ and $p_{fluct}$ are estimated from numerical simulations of \eqref{eq:virtual-fast-barotropic-QL} with parameters given in section \ref{sub:Numerical-implementation}, and $\alpha=0.073$. $p_{mean}$ is always positive, meaning that the average Reynolds' stresses are intensifying the zonal flow $U(\phi)$ at each latitude.
We see that fluctuations of Reynolds' stresses are lower at the jet
extrema ($p_{fluct}$ is relatively small), in particular close to the equator $\phi=0$. This can be understood as a consequence of the depletion of vorticity at the stationary streamline \cite{Bouchet_Morita_2010PhyD}. Error bars are not shown
here, see Figures \ref{fig:energy-balance-mean} and \ref{fig:energy-balance-fluct}.}

\end{center}\end{figure}

\begin{figure}\begin{center}
\subfigure[\label{fig:energy-balance-mean}]{\includegraphics[scale=0.5]{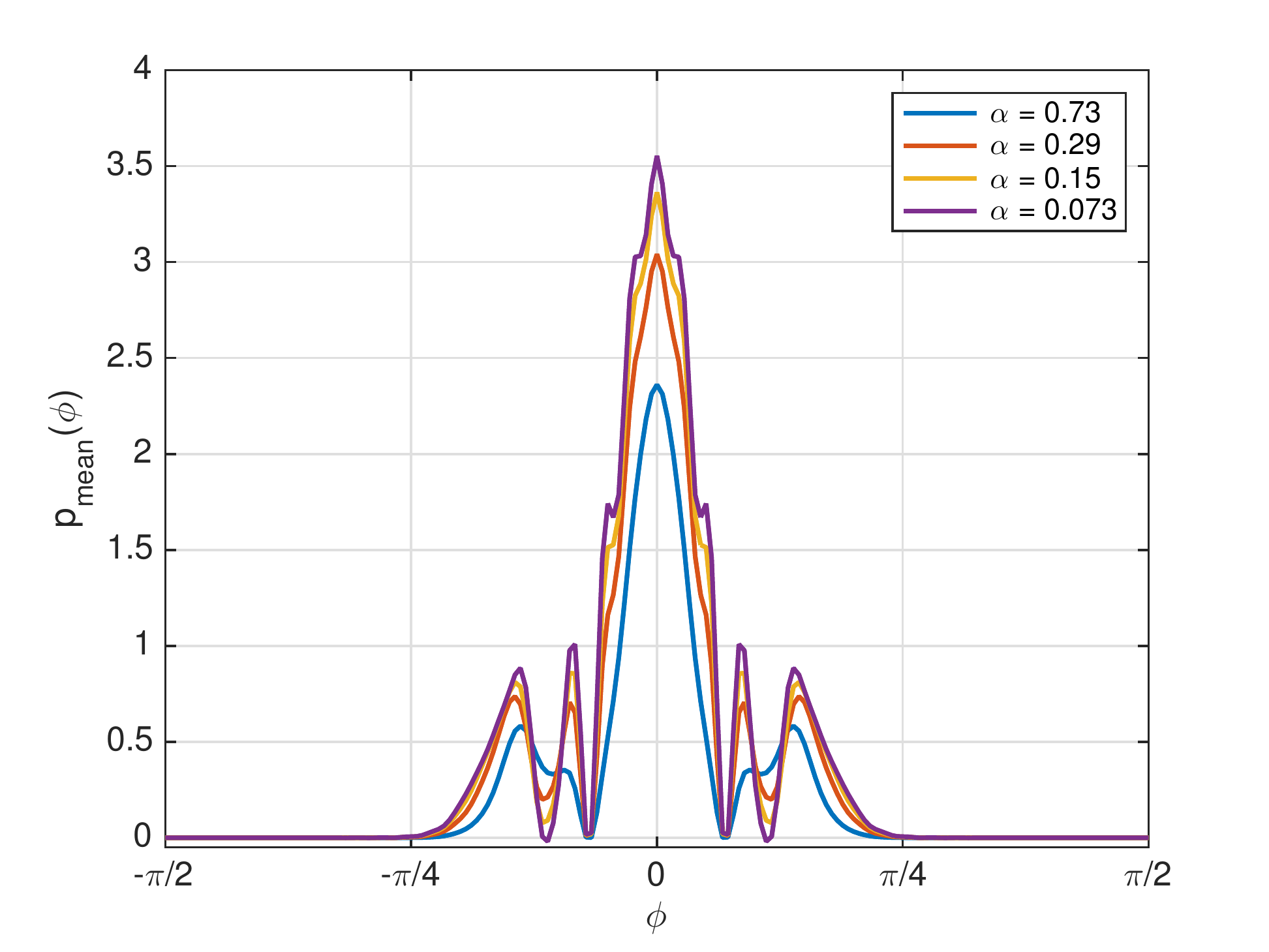}}\hspace{1cm}
\subfigure[\label{fig:energy-balance-fluct}]{\includegraphics[scale=0.5]{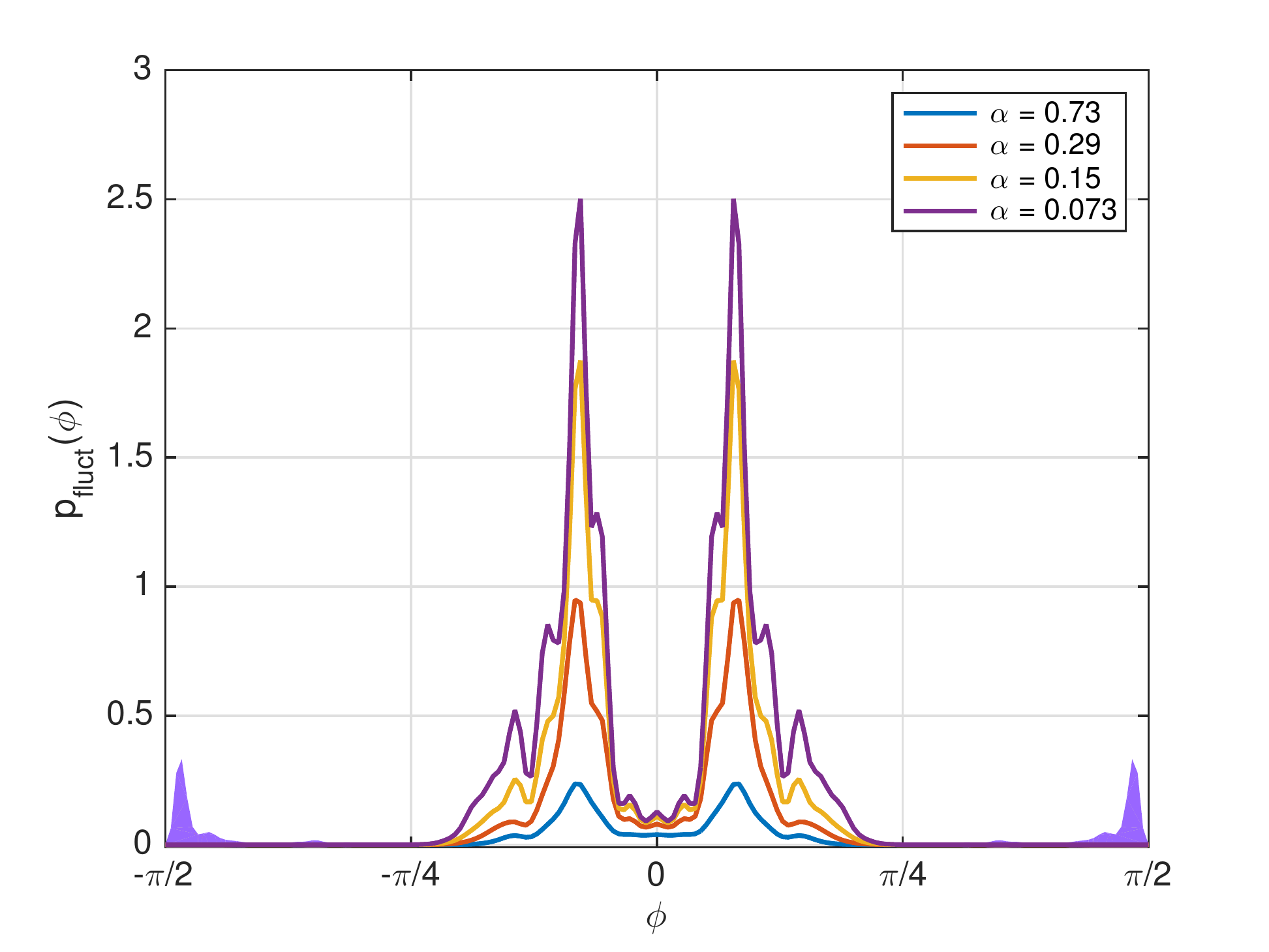}}
\caption{Energy injection rate into the zonal
flow (a) by the mean Reynolds' stresses $p_{mean}$ (first term in the
r.h.s of (\ref{eq:energy-balance-kinetic})) and (b) by the fluctuations of Reynolds' stresses $p_{fluct}$ (last
term in the r.h.s of (\ref{eq:energy-balance-kinetic})), as functions of latitude
$\phi$, estimated from direct numerical simulations of the linearized
barotropic equation \eqref{eq:virtual-fast-barotropic-QL} with parameters given in section \ref{sub:Numerical-implementation}, and with different
values of the damping rate $\alpha$. Shaded areas represent the statistical
error bars. In Figure (a), we observe the convergence of $p_{mean}$ to a finite function of $\phi$ as $\alpha\to0$, in agreement with the theoretical predictions. In Figure (b), we observe that the values of $p_{fluct}$ are relatively weak close the jet maximum $\phi=0$, while they keep increasing as $\alpha\to0$ in other locations, as expected from theory.}
\end{center}\end{figure}

The spatial distribution of the energy injection rates $p_{mean}(\phi)$ and $p_{fluct}(\phi)$ are represented in Figures \ref{fig:comparison-U-p} and \ref{fig:energy-balance-mean}, \ref{fig:energy-balance-fluct}. Both $p_{mean}(\phi)$ and $p_{fluct}(\phi)$ are concentrated in the jet region $\phi\in[-\pi/4,\pi/4]$, which is also the region where the stochastic forces act (see Figure \ref{fig:U}). 

In Figure \ref{fig:energy-balance-mean}, we observe that $p_{mean}$ is always positive. This means that the turbulent perturbations are everywhere injecting energy into the zonal degrees of freedom, i.e. the average Reynolds' stresses are intensifying the zonal flow $U(\phi)$ at each latitude. This effect is predominant at the jet maximum and around the jet minima (around $\phi=\pm\pi/8$). We also observe that $p_{mean}$ (and thus $F[U]$) converges to a finite value as $\alpha$ decreases. A similar result has been obtained for the two dimensional Navier--Stokes equation under the assumption that the linearized equation close to the base flow has no normal mode, using theoretical arguments \cite{Bouchet_Nardini_Tangarife_2013_Kinetic}. Those assumptions are not satisfied here, thus indicating that the finite limit of $F[U]$ as $\alpha$ vanishes is a more general result. This result is extremely important, indeed it implies that the effective dynamics \eqref{eq:LD-effective-U} is actually well-posed in the limit $\alpha\to0$.

By definition, $p_{fluct}(\phi)$ is necessarily positive. In Figure \ref{fig:energy-balance-fluct}, we see that $p_{fluct}(\phi)$ keeps increasing as $\alpha$ decreases in the region away from the jet maximum (roughly for $|\phi|\in[\pi/16,\pi/4]$). This is in contrast with the behaviour of $p_{mean}(\phi)$ (fig. \ref{fig:energy-balance-mean}). We note that such a behaviour for $p_{fluct}(\phi)$ has been obtained recently for the two-dimensional Navier-Stokes equation under the assumption that the base flow has no mode \cite{BouchetNardiniTangarife2015}. However, the range of values of $\alpha$ considered here is not wide enough to check precisely those theoretical results. 

We also observe in Figure \ref{fig:energy-balance-fluct} that $p_{fluct}(\phi)$ is relatively small in the region of jet maximum $\phi\simeq 0$. This means that Reynolds' stresses tend to fluctuate less in this area. In the context of the deterministic two-dimensional Euler equation linearized around a background shear flow, it is known that extrema of the background flow lead to a decay of the perturbation vorticity (depletion of the vorticity at the stationary streamline \cite{Bouchet_Morita_2010PhyD}). In a stochastic context, this implies that the perturbation vorticity $\delta\omega$ is expected to fluctuate less in the region of jet extrema, in qualitative agreement with what is observed in Figure \ref{fig:energy-balance-fluct}.

%
%

%%%%%%%%%%%%%%%%%%%%%%%%%%%%%%%%
\subsection{Empirical validation of the time scale separation hypothesis\label{sub:Empirical-validation-of}}

In this paper we assumed a large separation in time scales: the eddies $\delta\omega$ evolves much faster than the zonal flow $U$, permitting the quasilinear approximation. 
It has been shown in Ref. \onlinecite{Bouchet_Nardini_Tangarife_2013_Kinetic,tangarife-these} that for the linearized dynamics close to a zonal jet $U$, the autocorrelation function of both the eddy velocity and the Reynolds stresses are finite in the limit $\alpha\to0$, even if the dissipation vanishes in this limit. An effective dissipation  takes place, thanks to the Orr mechanism (see Refs. \onlinecite{Bouchet_Nardini_Tangarife_2013_Kinetic,tangarife-these}). This result ensures that time scale separation assumption is valid for small enough $\alpha$ (the eddies $\delta\omega$ evolve on a time scale of order one, and the zonal flow $U$ evolves on a time scale of order $1/\alpha$).

\begin{figure}\begin{center}
\includegraphics[scale=0.5]{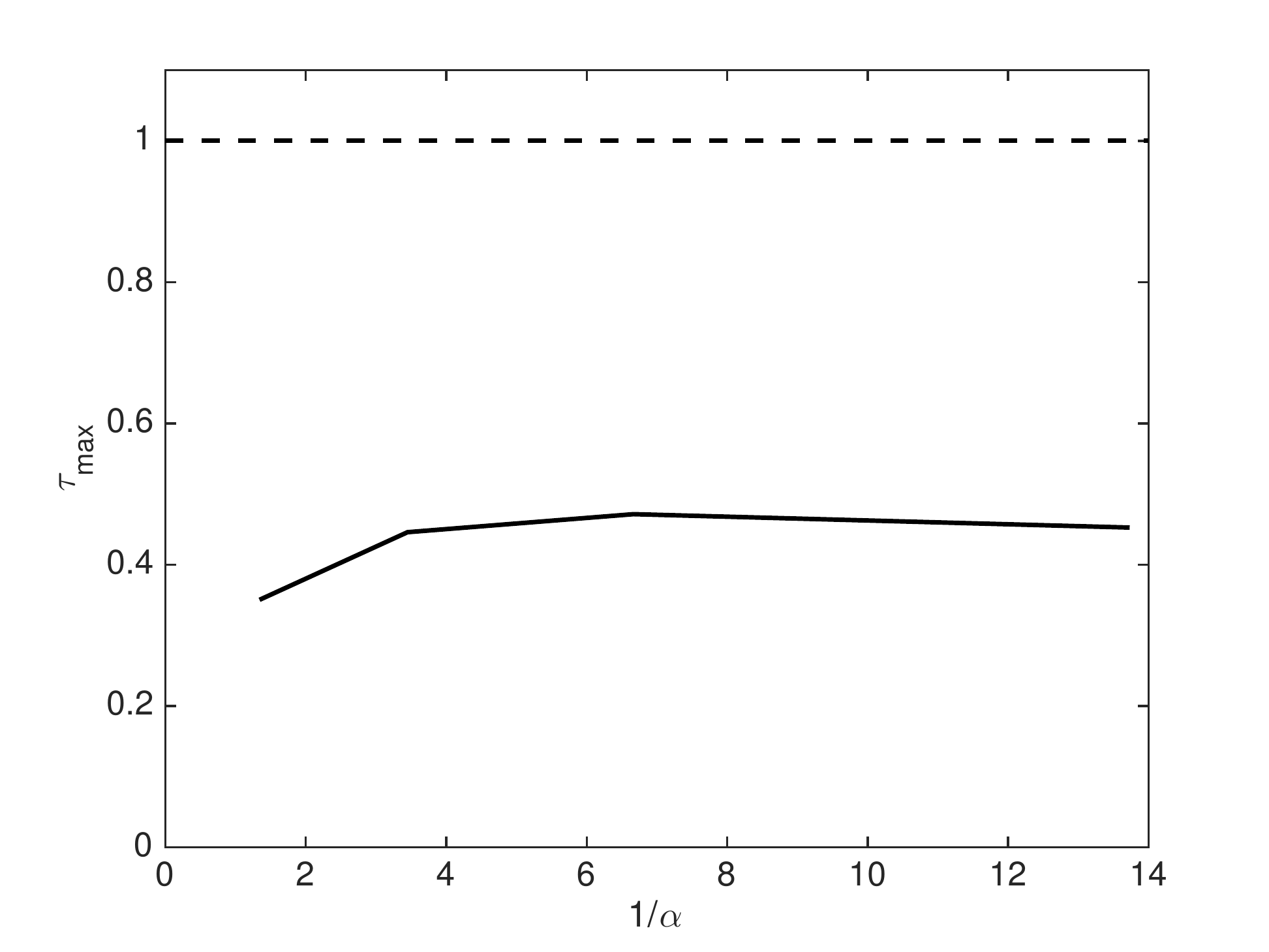}
\caption{\label{fig:autocorrelation-time-latitude}Solid line: maximum correlation time of
the Reynolds' stress divergence \eqref{eq:autocorrelation-time-phi}
as a function of the damping rate $\alpha$. We clearly see the convergence
of $\tau_{max}^\alpha$ to a finite value as $\alpha\to0$. The correlation time is of the order of the inertial time scale (equal to one by definition of the units, here represented by the dashed line), and much smaller than the
dissipative time $1/\alpha$ (not represented here), showing the time scale separation between
dissipative and inertial processes in the quasi-linear barotropic
dynamics.}
\end{center}\end{figure}

The consistency of this assumption for any value of $\alpha$ can also be tested numerically. For this purpose, we compute the maximum correlation
time of the Reynolds' stress divergence $f(\phi)$, defined as\footnote{In this spherical geometry the maximum is taken over the inner jet region $\phi\in[-\pi/7,\pi/7]$.}
\begin{equation}
\tau_{max}^\alpha\equiv\max_{\phi}\lim_{t\to\infty}\frac{1}{t}\int_{0}^{t}\int_{0}^{t}\frac{\mathbb{E}_{U}^\alpha\left[\left[f\left(\phi,s_{1}\right)f\left(\phi,s_{2}\right)\right]\right]}{2\mathbb{E}_{U}^\alpha\left[\left[f^{2}\left(\phi\right)\right]\right]}\,\mbox{d}s_{1}\mbox{d}s_{2}.\label{eq:autocorrelation-time-phi}
\end{equation}
We  check whether or not $\tau_{max}^\alpha \ll 1/\alpha$, where $1/\alpha$ is 
the dissipative time scale. The results are summarized in Figure \ref{fig:autocorrelation-time-latitude}.
We observe that $\tau_{max}^\alpha$ converges to a finite
value as $\alpha$ decreases, as expected from the theoretical analysis \cite{Bouchet_Nardini_Tangarife_2013_Kinetic,tangarife-these}, and this value is smaller than the inertial time scale (equal to one by definition of the time units). This means that the typical time scale
of evolution of the Reynolds' stress divergence is much smaller than the dissipative time scale $1/\alpha$ as soon as $1/\alpha$ is much larger than one, justifying the time scale separation hypothesis.

%%%%%%%%%%%%%%%%%%%%%%%%%%%%%%%%
%%%%%%%%%%%%%%%%%%%%%%%%%%%%%%%%
%%%%%%%%%%%%%%%%%%%%%%%%%%%%%%%%
\section{Large deviations of Reynolds stresses\label{sec:Large_Deviations_Reynolds_stresses}}
\label{sub:Application-of-the}

In section \ref{sub:Gaussian-approximation-of}, we studied the effective energy balance for the zonal flow $U(\phi)$ using numerical simulations of the linearized barotropic dynamics \eqref{eq:frozen-QL}. This effective description of zonal jet dynamics takes into account the low-order statistics of Reynolds' stresses: average and covariance.  In order to study rare events in zonal jet dynamics, we must employ the large deviation principle. The goal of this section is to apply the theoretical tools presented in sections \ref{sub:Large-time-large-deviations} and \ref{sec:LD-estimation-SCGF} to the study of rare events statistics in zonal jet dynamics.

%%%%%%%%%%%%%%%%%%%%%%%%%%%%%%%%
\subsection{Large Deviation Principle for the time-averaged Reynolds' stresses\label{sub:LD-LDP-time-average-stress}}

We first formulate the Large Deviation Principle for the quasi-linear barotropic equations \eqref{eq:barotropic-quasi-linear} in the regime $\alpha\ll1$, and present some properties of the large deviations functions. The numerical results are presented in section \ref{sec:LD-numerical-results}. The Large Deviation Principle presented here is equivalent to the one presented in a more general setting in section \ref{sub:Large-time-large-deviations}.

Consider the evolution of $\omega_z$ from the first equation of (\ref{eq:barotropic-quasi-linear}). Over a time scale $\Delta t$ much smaller than $1/\alpha$ but much larger than the correlation time $\tau$ we can write
\begin{equation}
\frac{\Delta \omega_z}{\Delta t} \equiv \frac{1}{\alpha}\frac{\omega_z(t+\Delta t) - \omega_z(t)}{\Delta t} \simeq \frac{1}{\Delta t}\int_t^{t+\Delta t} R(u)\,\mathrm{d} u - \omega_z(t)\,,
\label{eq:LD-omega_z-integral}
\end{equation}
where we have used the fact that $\omega_z$ has not evolved much from $t$ and $t+\Delta t$ (because $\Delta t\ll1/\alpha$), while $R(u)$ has evolved according to \eqref{eq:frozen-QL} with a fixed $\omega_z$ (or equivalently a fixed $U$). We also neglect hyper-viscosity in the evolution of $\omega_z$, which is natural in the turbulent
regime we are interested in. Note however that some hyper-viscosity is still present in the numerical simulations of (\ref{eq:frozen-QL}), in order to ensure numerical stability. For consistency, we make sure that the hyper-viscous terms have no influence on the numerical results (see Figure \ref{fig:H-section-5}).\\

We denote by $P_{\Delta t}\left[\frac{\Delta \omega_z}{\Delta t}\right]$ the probability distribution function of $\frac{\Delta \omega_z}{\Delta t}$, with a fixed $t$ (and thus a fixed $\omega_z(t)$), but with an increasing $\Delta t$.  This regime is consistent with the limit of time scale separation $\alpha\to0$, where $\omega_z$ is nearly frozen while $\delta\omega$ keeps evolving. From \eqref{eq:LD-omega_z-integral}, $P_{\Delta t}\left[\frac{\Delta \omega_z}{\Delta t}\right]$ is also the probability density function of the time-averaged advection term $\frac{1}{\Delta t}\int_t^{t+\Delta t} R(u)\,\mathrm{d} u$. The Large Deviation Principle gives the asymptotic expression of $P_{\Delta t}\left[\frac{\Delta \omega_z}{\Delta t}\right]$ in the regime $\Delta t \gg \tau$, namely
\begin{equation}
\ln P_{\Delta t}\left[\frac{\Delta \omega_z}{\Delta t}\right]\underset{\Delta t\to\infty}{\sim} -\Delta t\, \mathcal{L}\left[\frac{\Delta \omega_z}{\Delta t}\right]\,.
\label{eq:LD-LDP-deltaomega_z}
\end{equation}
The function $\mathcal{L}$ is called the large deviation rate function. It characterizes the whole distribution of $\frac{\Delta \omega_z}{\Delta t}$ in the regime $\Delta t\gg\tau$, including the most probable value and the typical fluctuations.\\

Our goal in the following is to compute numerically $\mathcal{L}\left[\frac{\Delta \omega_z}{\Delta t}\right]$. This can be done through the scaled cumulant generating function \eqref{eq:SCGF-general}. Using \eqref{eq:LD-omega_z-integral}, the definition \eqref{eq:SCGF-general} can be reformulated as
\begin{equation}
H[\theta] = \lim_{\Delta t\to\infty}\frac{1}{\Delta t}\ln \int \mathrm{d} \dot{\omega}_z \, P_{\Delta t}\left[\dot{\omega}_z\right] \exp\left(\theta\cdot \Delta t\, \dot{\omega}_z\right)
\label{eq:LD-SCGF-Gartner-Ellis}
\end{equation}
Because $\omega_z$ is a field, here $\theta$ is also a field depending on the latitude $\phi$, and $H$ is a functional. For simplicity, we stop denoting the dependency of $H$ in $\omega_z$. In \eqref{eq:LD-SCGF-Gartner-Ellis}, we also have used the notation
$\theta_{1}\cdot\theta_{2}\equiv\int\mbox{d}\phi\,\cos\phi\,\theta_{1}(\phi)\theta_{2}(\phi)$ for the canonical scalar product on the basis of spherical harmonics.

Using \eqref{eq:LD-LDP-deltaomega_z} in \eqref{eq:LD-SCGF-Gartner-Ellis} and using a saddle-point approximation to evaluate the integral in the limit $\Delta t\to\infty$, we get
$H[\theta] = \sup_{\dot{\omega}_z}\left\lbrace \theta\cdot\dot{\omega}_z  - \mathcal{L}\left[\dot{\omega}_z\right] \right\rbrace$, 
i.e. $H$ is the Legendre-Fenschel transform of $\mathcal{L}$. Assuming that $H$ is everywhere differentiable, we can invert this relation as
\begin{equation}
\mathcal{L}\left[\frac{\Delta \omega_z}{\Delta t}\right] = \sup_{\theta}\left\lbrace \theta\cdot\frac{\Delta \omega_z}{\Delta t}  - H[\theta] \right\rbrace\,.
\label{eq:LD-Legendre}
\end{equation}

The scaled cumulant generating function $H[\theta]$ can be computed either from a time series of $\delta\omega$ (see section \ref{sec:LD-estimation-SCGF}) or solving the Ricatti equation (see section \ref{sub:Quasi-linear-systems-with}). Then the large deviation rate function $\mathcal{L}$ can be computed using \eqref{eq:LD-Legendre}, and this gives the whole probability distribution of $\frac{\Delta \omega_z}{\Delta t}$ (or equivalently of the time-averaged Reynolds' stresses) through the Large Deviation Principle \eqref{eq:LD-LDP-deltaomega_z}.

In the following, we implement this program and discuss the physical consequences for zonal jet statistics. We first give a simpler expression of $H[\theta]$, that makes its numerical computation easier.

%%%%%%%%%%%%%%%%%%%%%%%%%%%%%%%%
\subsection{Decomposition of the scaled cumulant generating function}

Using the Fourier decomposition \eqref{eq:Fourier-definition}, we can decompose the perturbation vorticity as $\delta\omega(\lambda,\phi) = \sum_m \omega_m(\phi)\mathrm{e}^{im\lambda}$, where $\omega_m$ satisfies
\begin{equation}
\frac{\partial\omega_{m}}{\partial u}=-L_{U,m}\left[\omega_{m}\right]+\sqrt{2}\eta_{m},\label{eq:virtual-fast-barotropic-QL}
\end{equation}
where the Fourier transform of the linear
operator (\ref{eq:LD-linear-operator}) reads 
\begin{equation}
L_{U,m}\left[\omega_{m}\right](\phi)=-\frac{im}{\cos\phi}\left(U(\phi)\omega_{m}(\phi)+\gamma(\phi)\psi_{m}(\phi)\right)-\alpha\omega_{m}(\phi)-\nu_{n}\left(-\Delta_{m}\right)^{n}\omega_{m}(\phi).\label{eq:LD-linear-operator-m}
\end{equation}
In (\ref{eq:virtual-fast-barotropic-QL}), $\eta_{m}\left(\phi,t\right)$
is a Gaussian white noise such that $\eta_{-m}=\eta_{m}^{*}$, with
zero mean and with correlations 
\[
\mathbb{E}\left[\eta_{m}\left(\phi_{1},t_{1}\right)\eta_{m}^{*}\left(\phi_{2},t_{2}\right)\right]=c_{m}\left(\phi_{1},\phi_{2}\right)\delta(t_{1}-t_{2}),
\]
\[
\mathbb{E}\left[\eta_{m}\left(\phi_{1},t_{1}\right)\eta_{m}\left(\phi_{2},t_{2}\right)\right]=0,
\]
where $c_{m}$ is the $m$-th coefficient in the Fourier decomposition
of $C$ in the zonal direction.

Using the Fourier decomposition, the zonally averaged advection term can be written $R(\phi)=\sum_{m}R_{m}(\phi)$
with $R_{m}(\phi)=-\frac{im}{\cos\phi}\partial_{\phi}\left(\psi_{m}\cdot\omega_{-m}\right)$. Using this expression and the fact that $\omega_{m_{1}}$ and $\omega_{m_{2}}^{*}$
are statistically independent for $m_{1}\neq m_{2}$, the scaled cumulant generating function \eqref{eq:LD-SCGF-Gartner-Ellis} can be decomposed as\footnote{The time $t$ in the upper and lower bounds of the integral in \eqref{eq:SCGF-omegaz-H_m} are not relevant here, as we are considering the statistically stationary state of \eqref{eq:virtual-fast-barotropic-QL}.}
\begin{equation}\begin{aligned}
H[\theta] &\equiv \lim_{\Delta t\to\infty}\frac{1}{\Delta t}\ln \mathbb{E}_U\left[\exp\left(\theta\cdot \int_{0}^{\Delta t}\left(R(u)-\omega_{z}\right)\,\mbox{d}u \right)\right]\\
&=-\theta\cdot\omega_{z}+\sum_{m}H_{m}\left[\theta\right],\\
\end{aligned}\label{eq:SCGF-omegaz-H_m}\end{equation}
with
\begin{equation}
H_{m}\left[\theta\right]=\lim_{\Delta t\to\infty}\frac{1}{\Delta t}\log\mathbb{E}_{U}\exp\left[\int\mbox{d}\phi\,\cos\phi\,\theta\left(\phi\right)\int_{0}^{\Delta t}R_{m}\left(\phi,u\right)\,\mbox{d}u\right].
\label{eq:SCGF-QG-m}
\end{equation}
We recall that $\mathbb{E}_U$ is the average in the statistically stationary state of \eqref{eq:virtual-fast-barotropic-QL}.

In the following, we consider the case where only one Fourier
mode $m$ is forced, for simplicity and to highlight deviations from Gaussian statistics. If several modes are forced,
their contributions to the scaled cumulant generating function add
up, according to (\ref{eq:SCGF-omegaz-H_m}).\\

Finally, consider the decomposition of the zonally averaged advection term into
spherical harmonics \eqref{eq:spherical-harmonics-decomposition}, $R_m(\phi)=\sum_{\ell}R_{m,\ell}~P_{\ell}^{0}(\sin \phi)$.
Using $\theta(\phi)=\theta_{\ell} P_{\ell}^{0}(\sin \phi)$ in \eqref{eq:SCGF-QG-m},
we investigate the statistics of the $\ell$-th coefficient $R_{m,\ell}$.
The associated scaled cumulant generating function \eqref{eq:SCGF-QG-m}
is denoted $H_{m,\ell}\left(\theta\right)\equiv H_{m}\left[\theta P_{\ell}^{0}(\sin \phi)\right]$, and the large deviation rate function is denoted
\begin{equation}
\mathcal{L}_{m,\ell}\left(\dot{ \omega}_{\ell}\right) = \sup_{\theta_\ell}\left\lbrace \theta_\ell \,\dot{ \omega}_{\ell}  - H_{m,\ell}(\theta_\ell) \right\rbrace\,.
\label{eq:LD-Legendre-mell}
\end{equation}

%%%%%%%%%%%%%%%%%%%%%%%%%%%%%%%%
\subsection{Numerical results\label{sec:LD-numerical-results}}

The function $H_{m,\ell}$ defined in previous section can be computed either from a time series of $\omega_m(\phi,u)$ using the method described in section \ref{sec:LD-estimation-SCGF}, or solving the Ricatti equation as described in section \ref{sub:Quasi-linear-systems-with}. Then, the large deviation rate funtion is computed using \eqref{eq:LD-Legendre-mell}. We now show the results of these computations and discuss the physical consequences. We describe the results obtained by solving numerically the linearized barotropic equation \eqref{eq:frozen-QL}, where we use the mean flow $U$ the flow obtained from a quasilinear simulation as described in the end of section \ref{sub:Numerical-implementation}, and represented in Figure \ref{fig:U}.

\subsubsection{Scaled cumulant generating function}

\begin{figure}\begin{center}
\includegraphics[scale=0.5]{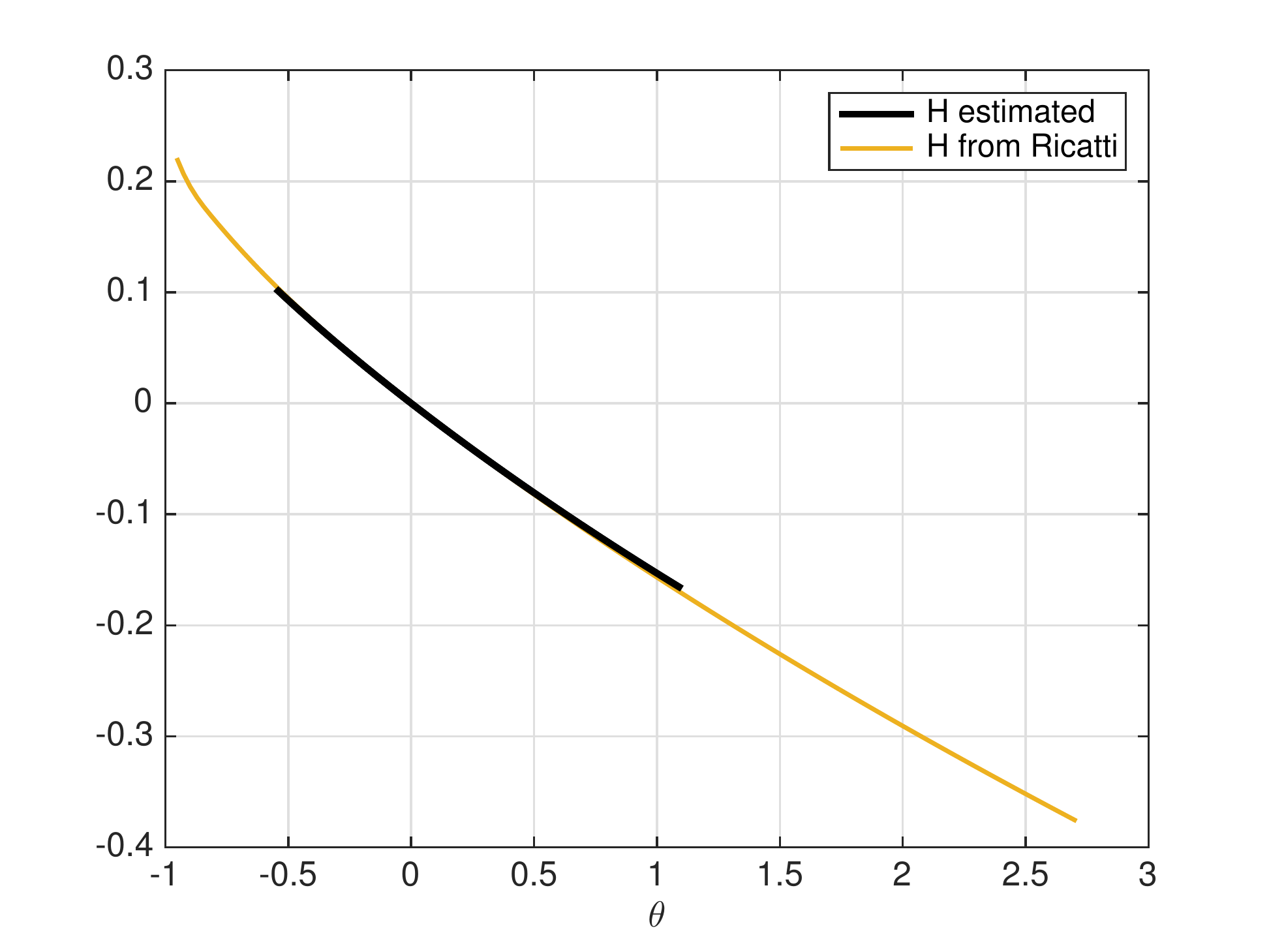}

\caption{\label{fig:H-section-5}Thick black line: scaled cumulant generating
function $H_{10,3}\left(\theta\right)$ estimated from the numerical
simulation of the linearized barotropic dynamics (\ref{eq:virtual-fast-barotropic-QL}),
with parameters defined in section \ref{sub:Numerical-implementation} and $\alpha=0.073$.
Statistical error bars are smaller than the width of this curve.
% Purple curve: quadratic fit that corresponds to the Gaussian approximation of $R_{10,3}(t)$, computed using the mean, variance and autocorrelation time of the time series of $R_{10,3}$.
Yellow curve:
scaled cumulant generating function $H_{10,3}\left(\theta\right)$
computed from numerical integration of the Ricatti equation (\ref{eq:NL-Lya}),
using (\ref{eq:SCGF-quasilinear}). The spectral cutoff in the Ricatti calculation is $L=120$ (compared to $L=80$ for the simulation of (\ref{eq:virtual-fast-barotropic-QL})), and the hyper-viscosity coefficient is such that the smallest scale has a damping rate of 4 (i.e. it is half of the hyperviscosity coefficient in the case $L=80$). The estimated scaled cumulant
generating function is in agreement with the one computed from the
Ricatti equation, showing that the finite spectral cutoff and hyperviscosity are negligible in the calculation of $H_{10,3}\left(\theta\right)$. The numerical integration
of the Ricatti equation enables access to larger values of $\left|\theta\right|$
(rarer events) extremely easily, see also Figure \ref{fig:L-section-5}.}
\end{center}\end{figure}

An example of computation of $H_{m,\ell}\left(\theta\right)$ is shown
in Figure \ref{fig:H-section-5}, with $m=10$, $\ell=3$ and $\alpha=0.073$.
The linearized barotropic equation (\ref{eq:virtual-fast-barotropic-QL})
is integrated over a time $T_{max}=54,500$, with fixed mean flow
given in Figure \ref{fig:U}, and the value of $R_{m,\ell}$ is recorded
every $0.03$ time units (the units are defined in section \ref{sec:LD-energy-balance-sphere}).

The scaled cumulant generating function (\ref{eq:SCGF-QG-m}) is
estimated following the procedure described in section \ref{sec:LD-estimation-SCGF}  (thick black curve in Figure \ref{fig:H-section-5}). Because the time series of $R_{m,\ell}$ is finite, $H_{m,\ell}(\theta)$ can only be computed with accuracy on a restricted range of values of $\theta$ (see section \ref{sub:Sampling-the-SCGF} for details), here $\theta\in[\theta_{min}/2,\theta_{max}/2] = [-0.6,1.1]$.\\

The scaled cumulant generating function (\ref{eq:SCGF-QG-m}) is also computed solving numerically
the Ricatti equation (\ref{eq:NL-Lya}) and using (\ref{eq:SCGF-quasilinear}) (yellow curve in Figure \ref{fig:H-section-5}). We observe almost perfect agreement between the direct estimation of $H_{m,\ell}$ (black curve in Figure \ref{fig:H-section-5}) and the computation of $H_{m,\ell}$ using the Ricatti equation (yellow curve). The integration of the Ricatti equation was done with a finer resolution and a lower hyper-viscosity than in the simulation of the linearized barotropic equation (\ref{eq:virtual-fast-barotropic-QL}), the agreement between both results in Figure \ref{fig:H-section-5} thus shows that the resolution used in the simulation of (\ref{eq:virtual-fast-barotropic-QL}) is high enough, and that the effect of hyper-viscosity is negligible.\\ 

We stress that the computation of $H_{m,\ell}(\theta)$ using the Ricatti equation (\ref{eq:NL-Lya}) does not require the numerical integration of the linear dynamics (\ref{eq:virtual-fast-barotropic-QL}). Typically, the integration of (\ref{eq:virtual-fast-barotropic-QL}) over a time $T_{max}=54,500$ takes about one week, while the resolution of the Ricatti equation (\ref{eq:NL-Lya}) for a given value of $\theta$ is a matter of a few seconds. This enables the investigation of the statistics of rare events (large values of $\left|\theta\right|$ in Figure \ref{fig:H-section-5}) extremely
easily, as we now explain in more detail.

\subsubsection{Rate function and departure from Gaussian statistics}

The main goal of this study is to investigate the statistics of rare events in zonal jet dynamics, that cannot be described by the effective dynamics studied in section~\ref{sub:Gaussian-approximation-of}. Using the previous numerical results, we now show how to quantify the departure from the effective description.

%We stress that the average dynamics of zonal jets in the regime $\alpha\ll1$ is fully described by the kinetic equation \eqref{eq:LD-effective-U}. As a consequence, the numerical results shown in section \ref{sub:Gaussian-approximation-of} summarize how the low-order statistics of the time-averaged advection term depend on the physical parameters $\alpha$ and $\nu_n$ (see for instance Figure \ref{fig:energy-balance-total}). For this reason, we will focus here on the probabilities of very rare events, as described by the Large Deviation Principle \eqref{eq:LD-LDP-deltaomega_z}.\\

\begin{figure}\begin{center}
%\subfigure[$\alpha=0.073$]{\includegraphics[scale=0.5]{./5LargeDev/L_0073}}
%\subfigure[$\alpha=0.73$]{\includegraphics[scale=0.5]{./5LargeDev/L_073}}
\includegraphics[scale=0.5]{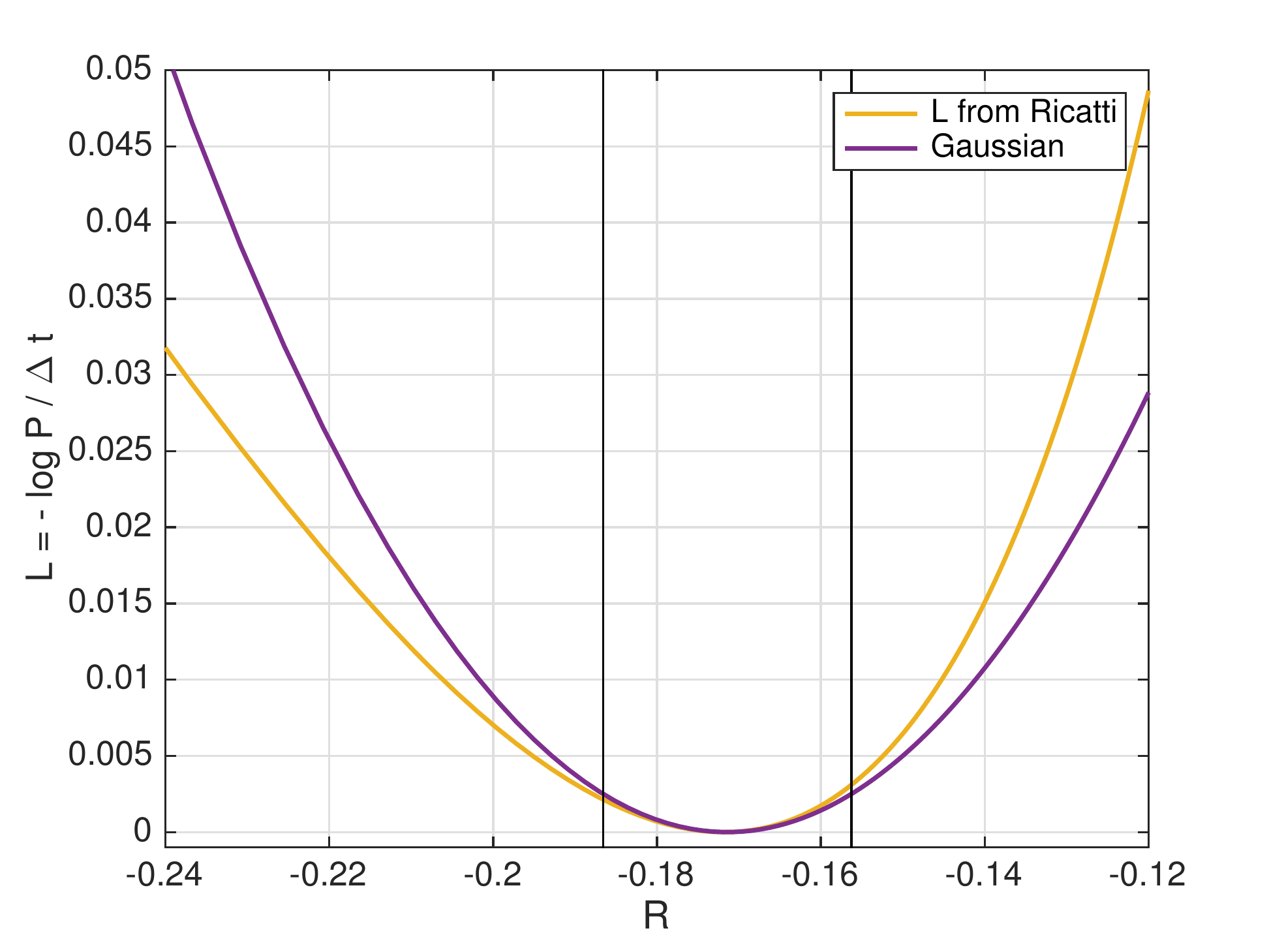}
\caption{\label{fig:L-section-5}Yellow curve: large deviation rate
function $\mathcal{L}_{10,3}(\bar{R})$ computed from numerical integration of the Ricatti equation (\ref{eq:NL-Lya}),
using (\ref{eq:SCGF-quasilinear}) and \eqref{eq:LD-Legendre}, with parameters defined in section \ref{sub:Numerical-implementation} and $\alpha=0.073$. Purple curve: quadratic fit \eqref{eq:LD-expansion-Lmell}
that corresponds to a Gaussian process with same mean and variance as $\bar{R}_{10,3,\Delta t}$, the time-averaged advection term. Black vertical lines: standard deviation of $\bar{R}_{10,3,\Delta t}$ around its mean. Outside the vertical lines, we observe non-Gaussian behaviour of $\bar{R}_{10,3,\Delta t}$, in particular negative fluctuations are much more probable than positive ones.}
\end{center}\end{figure}

The large deviation rate function $\mathcal{L}_{m,\ell}$ entering in the Large Deviation Principle \eqref{eq:LD-LDP-deltaomega_z} can be computed from $H_{m,\ell}$ using \eqref{eq:LD-Legendre-mell}. The result of this calculation\footnote{Here the Legendre-Fenschel transform \eqref{eq:LD-Legendre} is estimated as $\mathcal{L}_{m,\ell}(\dot{\omega}_z)=\theta^\star \cdot \dot{\omega}_z - H_{m,\ell}\left(\theta^\star\right) $ where $\theta^\star$ is the solution of $\dot{\omega}_z = \partial_\theta H_{m,\ell}\left(\theta^\star\right) $. Other estimators could be considered \cite{rohwer2014convergence}.} is shown in Figure \ref{fig:L-section-5} (yellow curve).\\
Because of the relation \eqref{eq:LD-omega_z-integral}, $\mathcal{L}_{m,\ell}$ can also be interpreted as the large deviation rate function for the time-averaged advection term, denoted $\bar{R}_{m,\ell,\Delta t}\equiv \frac{1}{\Delta t}\int_0^{\Delta t}R_{m,\ell}(u)\,\mathrm{d} u$. In other words, the probability distribution function of $\bar{R}_{m,\ell,\Delta t}$ in the regime $\Delta t\gg\tau$ satisfies
\begin{equation}
\ln P_{m,\ell,\Delta t}\left(\bar{R}\right) \underset{\Delta t\gg\tau}{\sim} -\Delta t \,\mathcal{L}_{m,\ell}\left(\bar{R}\right).
\label{eq:LD-LDP-Lmell}
\end{equation}

The Central Limit Theorem states that for large $\Delta t\gg\tau$, the statistics of $\bar{R}_{m,\ell,\Delta t}$ around its mean $\mathcal{R}_{m,\ell}\equiv\mathbb{E}_U\left[\bar{R}_{m,\ell,\Delta t}\right]=\mathbb{E}_U\left[R_{m,\ell}\right]$ are nearly Gaussian. A classical result in Large Deviation Theory is that the Central Limit Theorem can be recovered from the Large Deviation Principle \cite{freidlin2012random}. Indeed, using the Taylor expansion of $H_{m,\ell}$ in powers of $\theta$ (\ref{eq:SCGF-expansion-theta}) and computing the Legendre-Fenschel transform \eqref{eq:LD-Legendre}, we get
\begin{equation}
\mathcal{L}_{m,\ell}\left(\bar{R}\right)  = \frac{1}{2\mathcal{Z}_{m,\ell}}\left(\bar{R} - \mathcal{R}_{m,\ell}\right)^2 + O\left(\left(\bar{R} - \mathcal{R}_{m,\ell}\right)^3\right)
\label{eq:LD-expansion-Lmell}
\end{equation}
with $\mathcal{Z}_{m,\ell}\equiv \lim_{\Delta t\to\infty}\Delta t\,\mathbb{E}_{U}\left[\left[\bar{R}_{m,\ell,\Delta t}^2\right]\right]$. Using the Large Deviation Principle \eqref{eq:LD-LDP-Lmell}, this means that the statistics of $\bar{R}_{m,\ell,\Delta t}$ for small fluctuations around $\mathcal{R}_{m,\ell}$ are Gaussian with variance $\mathcal{Z}_{m,\ell}/\Delta t$, which is exactly the result of the Central Limit Theorem. Then, the difference between the actual rate function $\mathcal{L}_{m,\ell}\left(\bar{R}\right)$ and its quadratic approximation (right-hand side of \eqref{eq:LD-expansion-Lmell}) gives the departure from the Gaussian behaviour of $\bar{R}_{m,\ell,\Delta t}$.\\

From \eqref{eq:LD-expansion-Lmell}, the Gaussian behaviour is expected to apply roughly for $\left|\bar{R} - \mathcal{R}_{m,\ell}\right|\leq \sigma_{m,\ell,\Delta t}$ with $\sigma_{m,\ell,\Delta t} \equiv \sqrt{\mathcal{Z}_{m,\ell}/\Delta t}$. The values of $\mathcal{R}_{m,\ell}\pm\sigma_{m,\ell,\Delta t}$ are represented by the black vertical lines in Figure~\ref{fig:L-section-5}\footnote{The value of $\Delta t$ used in this estimation is the optimal one $\Delta t^\star$, defined in section \ref{sec:LD-estimation-SCGF}.}. The quadratic approximation of the rate function is also shown in Figure \ref{fig:L-section-5} (purple curve). As expected, the curves are indistinguishable from each other between the vertical lines (typical fluctuations), and departures from the Gaussian behaviour are observed away from the vertical lines (rare fluctuations). Namely, the probability of a large negative fluctuation is much larger than the probability of an equally large fluctuation for a Gaussian process with same mean and variance as $\bar{R}_{m,\ell,\Delta t}$. On the contrary, the probability of a large positive fluctuation is much smaller than the the probability of the same fluctuation for a Gaussian process with same mean and variance as $\bar{R}_{m,\ell,\Delta t}$.\\

The kinetic description basically amounts at replacing $\bar{R}_{m,\ell,\Delta t}$ by a Gaussian process with same mean and variance. From the results summarized in Figure \ref{fig:L-section-5}, we see that such approximation leads to a very inaccurate description of rare events statistics. Understanding the influence of the non-Gaussian behavior of $\bar{R}_{m,\ell,\Delta t}$ on zonal jet dynamics is naturally a very interesting perspective of this work.

%Now consider a fluctuation of $\bar{R}_{m,\ell,\Delta t}$ from its mean by an amount $A\times\sigma_{m,\ell,\Delta t}$. A natural question to ask is: what is the ratio between the probability of this fluctuation estimated from the Large Deviation Principle on one hand and estimated from the Central Limit Theorem on the other hand? We expect this difference to be small for $A\leq1$ (typical fluctuations, between the vertical lines in Figure \ref{fig:L-section-5}) and to be large for $A\geq1$ (rare fluctuations, outside the vertical lines in Figure \ref{fig:L-section-5}).

\section{Conclusions and perspectives\label{conclusions_perspectives}}

In this work we carried out a first study of the typical and large fluctuations of the Reynolds stress in fluid mechanics. Reynolds stress is certainly a key quantity in studying the largest scales of turbulent flows. This is especially true whenever a time scale separation is present, in which case it can be expected that an effective slow equation governs the large scale flow evolution (see equation \eqref{eq:Slow_Stochastic_Dynamics}). Not only the averaged momentum flux (the Reynolds stress) and averaged advection terms are essential, but also their fluctuations (that we call the Reynolds stress fluctuations). 

We studied the case of a zonal jet for the barotropic equation on a sphere, in a regime for which time scale separation is relevant. For this case, we show that the probability distribution function of the equal-time (without time average) advection term has a distribution with typical fluctuations which are very large compared to the average, and with heavy tails. These probability distribution functions have exponential tails, both for the quasilinear and fully non-linear dynamics cases.  For quasilinear dynamics we gave a simple explanation for these exponential tails.

When one is interested in the low frequency evolution of the jet, these high frequency fluctuations of the advection term and momentum fluxes are not relevant. We discussed that the natural quantity to study is the large deviation rate function for the time averaged advection term (that we call the Reynolds stress large deviation rate function). We have proposed two methods to compute this rate function. First an empirical method, directly from the time series of the advection term, that could be applied to any dynamics. Second we show that for the quasilinear dynamics, the Reynolds stress large deviation rate function can be computed as the contraction of a solution of a matrix Riccati equation. We demonstrated that such a computation can be performed by generalizing classical algorithms used to compute Lyapunov equations. Solving the matrix Riccati equation is much more computationally efficient, by several orders of magnitude, compared to accumulating statistics by numerical simulation, and gives direct and easy access to the probability of rare events.   The approach is however limited to the quasilinear dynamics so far. 

We discussed the Reynolds stress large deviation rate again for the specific case of a zonal jet that arises in turbulent barotropic flow on the rotating sphere. We illustrated the computation of the Reynolds stress large deviation rate, both using the empirical method and the Riccati equation. These two approaches give a very good agreement. This large deviation rate function clearly illustrate the existence of non-Gaussian fluctuations. The non Gaussian fluctuations are much more rare than Gaussian ones for positive values of the Reynolds stress component and much less rare than Gaussian for negative values. 

Our work illustrates the possibility to compute Reynolds stress large deviation rate functions. It opens up a number of perspectives. A next step would be to study the spatial structure of the Reynolds stress fluctuation, and describe it from a fluid mechanics perspective.  It would help to answer the following questions: What are the dominant spatial pattern for the fluctuations of the Reynolds stresses? What causes them? What is their effect on the low frequency variability of the large scale flow? 
The most interesting application of the Reynolds stress large deviation rate functions may be the study of rare long term evolutions of the large scale flow. For instance, in many examples, rare transitions between turbulent attractors have been observed, leading to a bistability phenomenology. In order to study quantitatively such a bistability phenomenology, for instance in order to compute transitions rates and transitions paths between attractors, one could consider equation (\ref{eq:Slow_Stochastic_Dynamics}) in the framework of Freidlin--Wentzell theory. The large deviation rate function we studied in this work would then be the basic building block, that would allow to define an action that should be minimized to compute transition paths and transition rates. In order to compute the action, the large deviation rate function should then be computed for any flow $U$ along a possible transition path, as described in section \ref{sec:LD-numerical-results} for a single example of a flow $U$.

An essential question, at a more mathematical level, is the validity of the quasilinear approximation as far as rare events are concerned. The self consistency of the quasilinear approach has been discussed theoretically by focusing on the average Reynolds stress \cite{Bouchet_Nardini_Tangarife_2013_Kinetic}. This point has also been verified numerically in this work, through the study of properties of the energy balance (see section \ref{sub:energy-balance}) and through the verification of the fact that the linear equation correlation time has a limit when $\alpha \rightarrow 0$ (see section \ref{sub:Empirical-validation-of}). However this does not necessarily imply that the quasilinear approximation is self-consistent as far as fluctuations, and more specifically rare fluctuations, are concerned. This could be addressed by studying the properties of solutions to the Ricatti equation in the limit $\alpha\to0$ to assess whether or not the small scale dissipative mechanism (either viscosity or hyperviscosity) affects the statistics of the rare fluctuations.  This problem is left as a prospect for future work.

\begin{acknowledgements} 
The research leading to these results has received funding from the European Research Council under the European Union's Seventh Framework Programme (FP7/2007-2013 Grant Agreement No. 616811) (F. Bouchet and T. Tangarife) and from the US NSF under Grant No. DMR-1306806 (J. B. Marston). J. B. Marston would also like to thank the Laboratoire de Physique de l'ENS de Lyon and CNRS for hosting a visit where some of this work was carried out. We thank the reviewers for their extremely careful reading of our paper and for their useful suggestions. 
\end{acknowledgements}

\bibliography{biblio}

\end{document}